\documentclass[sigconf]{acmart}
\AtBeginDocument{%
  }

\setcopyright{acmlicensed}
\copyrightyear{2018}
\acmYear{2018}
\acmDOI{XXXXXXX.XXXXXXX}
\acmConference[Conference acronym 'XX]{Make sure to enter the correct
  conference title from your rights confirmation email}{June 03--05,
  2018}{Woodstock, NY}
\acmISBN{978-1-4503-XXXX-X/2018/06}

\usepackage{tikz}
\usepackage{amsmath}
\usepackage{color}
\usepackage{filecontents}

\usepackage[linesnumbered,boxed,ruled,commentsnumbered]{algorithm2e}
\usepackage{amsmath}
\usepackage{comment}
\usepackage{hyperref} 
\usepackage{cleveref}
\usepackage{amssymb}
\usepackage{graphicx}
\usepackage{subfigure}
\usepackage{algorithmic}
\usepackage{xcolor}
\usepackage{url}
\usepackage{xspace}
\usepackage{balance}
\usepackage{colortbl} 
\usepackage{booktabs}

\usepackage{enumitem}
\usepackage{arydshln}
\usepackage{caption}
\usepackage{bookmark}
\setlist[itemize]{leftmargin=*}
\newcommand{\norm}[1]{\left\lVert#1\right\rVert}
\newtheorem{example}{Example}
\newtheorem{theorem}{Theorem}
\newtheorem{lemma}{Lemma}
\newtheorem{definition}{Definition}

\newtheorem{remark}{Remark}

\hypersetup{
  colorlinks=true,
  linkcolor={blue!70!green},
  citecolor={green!70!blue},
  urlcolor={orange!70!red}
}

\newcommand{\twopartdef}[4]
{
	\left\{
	\begin{array}{ll}
		#1 & #2 \\
		#3 & #4
	\end{array}
	\right.
}

\newcommand{\mypara}[1]{\smallskip\noindent\textbf{#1.}\xspace}

\newcommand{\method}{\textsf{MaskedXGBoost}\xspace}

\DeclareGraphicsExtensions{%
    .png,.PNG,%
    .pdf,.PDF,%
    .jpg,.mps,.jpeg,.jbig2,.jb2,.JPG,.JPEG,.JBIG2,.JB2}

\begin{document}

\title{Bilateral Differentially Private Vertical Federated Boosted Decision Trees}



\author{Bokang Zhang}
\affiliation{%
  \institution{The Chinese University of Hong Kong, Shenzhen}
  \city{}
  \country{}}
\email{bokangzhang@link.cuhk.edu.cn}

\author{Zhikun Zhang}
\affiliation{%
  \institution{Zhejiang University}
  \city{}
  \country{}}
\email{zhikun@zju.edu.cn}

\author{Haodong Jiang}
\affiliation{%
  \institution{The Chinese University of Hong Kong, Shenzhen}
  \city{}
  \country{}}
\email{haodongjiang@link.cuhk.edu.cn}

\author{Yang Liu}
\affiliation{%
  \institution{Bytedance Inc.}
  \city{}
  \country{}}
\email{liuyang.fromthu@bytedance.com}

\author{Lihao Zheng}
\affiliation{%
  \institution{The Chinese University of Hong Kong, Shenzhen}
  \city{}
  \country{}}
\email{lihaozheng@link.cuhk.edu.cn}

\author{Yuxiao Zhou}
\affiliation{%
  \institution{The Chinese University of Hong Kong, Shenzhen}
  \city{}
  \country{}}
\email{yuxiaozhou@link.cuhk.edu.cn}

\author{Shuaiting Huang}
\affiliation{%
  \institution{Zhejiang University}
  \city{}
  \country{}}
\email{shuait_huang@zju.edu.cn}

\author{Junfeng Wu}
\affiliation{%
  \institution{The Chinese University of Hong Kong, Shenzhen}
  \city{}
  \country{}}
\email{junfengwu@cuhk.edu.cn}

\renewcommand{\shortauthors}{Zhang et al.}

\begin{abstract}
Federated learning is a distributed machine learning paradigm that enables collaborative training across multiple parties while ensuring data privacy. 
Gradient Boosting Decision Trees (GBDT), such as XGBoost, have gained popularity due to their high performance and strong interpretability.
Therefore, there has been a growing interest in adapting XGBoost for use in federated settings via cryptographic techniques. 
However, it should be noted that these approaches may not always provide rigorous theoretical privacy guarantees and they often come with a high computational cost in terms of time and space requirements.
In this paper, we propose a variant of vertical federated XGBoost with bilateral differential privacy guarantee: \method.
We build well-calibrated noise to perturb the intermediate information to protect privacy.
The noise is structured with part of its ingredients in the null space of the arithmetical operation for splitting score evaluation in XGBoost, helping us achieve consistently better utility than other perturbation methods and relatively lower overhead than encryption-based techniques. 
We provide theoretical utility analysis and empirically verify privacy preservation.
Compared with other algorithms, our algorithm's superiority in both utility and efficiency has been validated on multiple datasets.
\end{abstract}

\maketitle

\section{Introduction}\label{sec:intro}
Big data is widely recognized to play an essential role in machine learning~\cite{yang2019federated}.
Traditionally, large datasets are aggregated from multiple sources and processed by a central server through a process known as \textit{centralized learning}.
However, this paradigm is becoming increasingly problematic due to concerns over the unauthorized use and exploitation of users' personal data.
\textit{Federated learning} (FL)~\cite{2017googlefl} addresses this challenge by transferring intermediate results of the training algorithm instead of the raw personal data.
Based on how data is partitioned, FL can be roughly classified into two categories: Horizontal FL (HFL) and vertical FL (VFL). 
HFL, also known as sample-wise FL, targets the scenarios where participants' data have the same feature space but differ in samples~\cite{yang2019federated}. 
VFL, or feature-wise FL, is applicable to cases where multiple datasets share the same sample ID space but differ in feature space~\cite{yang2019federated}.
While HFL has been extensively studied by the research community~\cite{shokri2015privacy,mcmahan2016federated, bonawitz2017practical,aono2017privacy, maddock2022hltreewithdp}, less attention has been given to VFL. 
Existing VFL methods heavily rely on cryptography technologies such as homomorphic encryption and secure multiparty computation to calculate privacy-related intermediate results~\cite{cheng2021secureboost, yang2019federated}.

Gradient Boosting Decision Trees (GBDT), such as XGBoost~\cite{chen2016}, are well-regarded for their high performance and interoperability.
They have found widespread use in various industrial applications, including financial risk management \cite{lu2023squirrel,kagglecredit,tian2020credit} and online advertising \cite{gohil2021click,ling2017model}.
In these applications, training data and labels often belong to separate parties. 
For example, e-commerce companies may want to personalize product recommendations using financial data, like users' bank loans. 
In this scenario, label data is with the e-commerce company, and financial data is at the bank. 
This setup falls under VFL.
Our paper focuses on extending the benefits of XGBoost to VFL settings with privacy guarantees.

\mypara{Existing Solutions}
Most of the existing studies on VFL-setting XGBoost require different types of encryption-based protocols: Homomorphic encryption (HE)~\cite{cheng2021secureboost,jin2022oppoboost, lu2023squirrel} and secret-sharing~\cite{fang2020hybrid}, resulting in a large communication and computing overhead. 
Another line of research adopts the notion of \textit{differential privacy} (DP) or \textit{local differential privacy} (LDP) and performs the analysis on the perturbed data~\cite{li2020privacy, tian2023sf, wang2022feverless}.
Le et al.~\cite{le2021fedxgboost} propose perturbing the gradient information at the label party, while Tian et al.~\cite{tian2023sf} propose perturbing the feature information on the feature party side.
Despite DP-based methods reducing the training time a lot, the model suffers accuracy loss from the injected noise.
The work~\cite{le2021fedxgboost} also proposes to adopt \textit{secure matrix multiplication}~\cite{karr2009privacy} in the private training of vertical federated XGBoost~\cite{le2021fedxgboost}.
Secure matrix multiplication is a cryptographic technique that allows two parties to jointly compute the product of their matrices without revealing their inputs to each other.
This approach offers low communication costs but without rigorous proof of privacy guarantee.

\mypara{Our Contributions}
Our proposed \method ensures bilateral differential privacy guarantees for both parties during the training, and our method achieves consistently better model utility.
We consider the evaluation of a splitting score, the major step requiring privacy guarantee in XGBoost, as the multiplication of a categorical matrix and a vector~\cite{le2021fedxgboost}. 
Our approach constructs well-calibrated noises to the sensitive information vector, which significantly enhances privacy while having a minimal impact on the utility of the data. 
Unlike previous research that only focused on analyzing the privacy leakage of one party holding the label, our approach employs differential privacy to ensure privacy preservation for both parties involved. 
Additionally, we build an attack model, trying to reveal any sensitive label information of users to check whether our protocol is secure enough to protect against label inference attacks.
Our main contributions are as follows:
\begin{itemize}
    \item \mypara{(C1) \method}
    We propose an algebra-based approach, called \method, to achieve better model utility than perturbation methods and higher efficiency than HE methods.
    We conduct extensive experiments on six datasets to illustrate the improvement of utility and efficiency of \method.
    It achieves $4.82 \times$ (Adult dataset) to $6.72 \times$ (Nomao dataset) training time improvement.
    We provide a theoretical analysis of the better utility property of \method in \autoref{theorem:utility}.

    \item \mypara{(C2) New Idea for Designing Differential Private Mechanisms with Better Utility}
    Rather than relying on conventional Gaussian mechanisms or other common differential privacy noise mechanisms, our approach involves designing noises tailored to the specific problem being solved to achieve differential privacy. 
    Most of the noise is allocated to the null space of the arithmetical operation for splitting score evaluation in XGBoost, striking an acceptable balance between training utility and data privacy.

    \item \mypara{(C3) Differential Privacy Analysis for Both Parties} 
    We provide sufficient differential privacy proof for the so-called active party and passive party in the VFL setting. 
    To the best of our knowledge, our bilateral privacy analysis is the first in the literature. 
    In particular, we discover that different noise ingredients (see \autoref{subsec:noise_calibration} for greater detail of noise calibration) affect the two parties of \method differently. 
    The energy of the whole noise should be considered for the active party's privacy protection(see \autoref{theorem:APLDP}), while the energy ratio of different ingredients and the size of a training dataset are the main factors for the passive party's privacy (see \autoref{theorem:PPDP}).
    We conduct the ablation study of the effect of the noise energy ratio on utility.
    
    \item \mypara{(C4) Empirical Privacy Evaluation} 
    We conduct empirical privacy evaluation by building a label inference attacker and an attribute inference attacker to verify the theoretical differential privacy results(see \autoref{theorem:APLDP} and \autoref{theorem:PPDP}). 
    Under a strict privacy budget, the attacker's performance is poor(i.e., attack acc $0.51$ for AP and $0.52$ for PP), demonstrating the security and privacy of our approach.
\end{itemize}

\section{Preliminaries} 
\label{sec:Preliminaries}

\subsection{Vertical Federated Learning}
\label{subsec:pre_vfl}

\textit{Federated learning} (FL) aims to facilitate the collaboration of multiple participants in contributing distinct training data to train a better model collectively. Based on how the data is partitioned, FL can be roughly classified into two categories~\cite{yang2019federated}: \textit{Horizontal federated learning} (HFL) and \textit{vertical federated learning} (VFL). 
In this paper, we focus on VFL.

\mypara{VFL}
Vertical FL, also known as feature-wise FL, is suitable for situations where multiple datasets share the same sample ID space but have different feature spaces.
For example, consider two different companies in a city, one is a bank, and the other is an e-commerce company. 
Their user sets are likely to contain most of the residents of the area, so the intersection of their user spaces is large. 
Since the bank has access to the user's financial information, such as revenue and expenditure behavior, while the e-commerce company retains the user's browsing and purchasing history, their feature spaces are significantly distinct. 
We want both parties to have a prediction model for product purchase based on user and product information.

\subsection{XGBoost}

Consider a dataset $\mathcal{D}$ = $\{(x_i, y_i), x_i \in \mathbb R^d, y_i \in \mathbb R, i\in\mathcal I\}$, where $x_i$ denotes the feature vectors of the $i^{th}$ instance in a feature space $\mathcal{X}$, $y_i$ is the label of the $i^{th}$ instance, and $\mathcal I$ denotes the index set of instances. 
Let $n = |\mathcal{I}|$ be the total number of instances. 

\begin{figure} [!tbp]
\centering  
\includegraphics[width=0.4\textwidth]{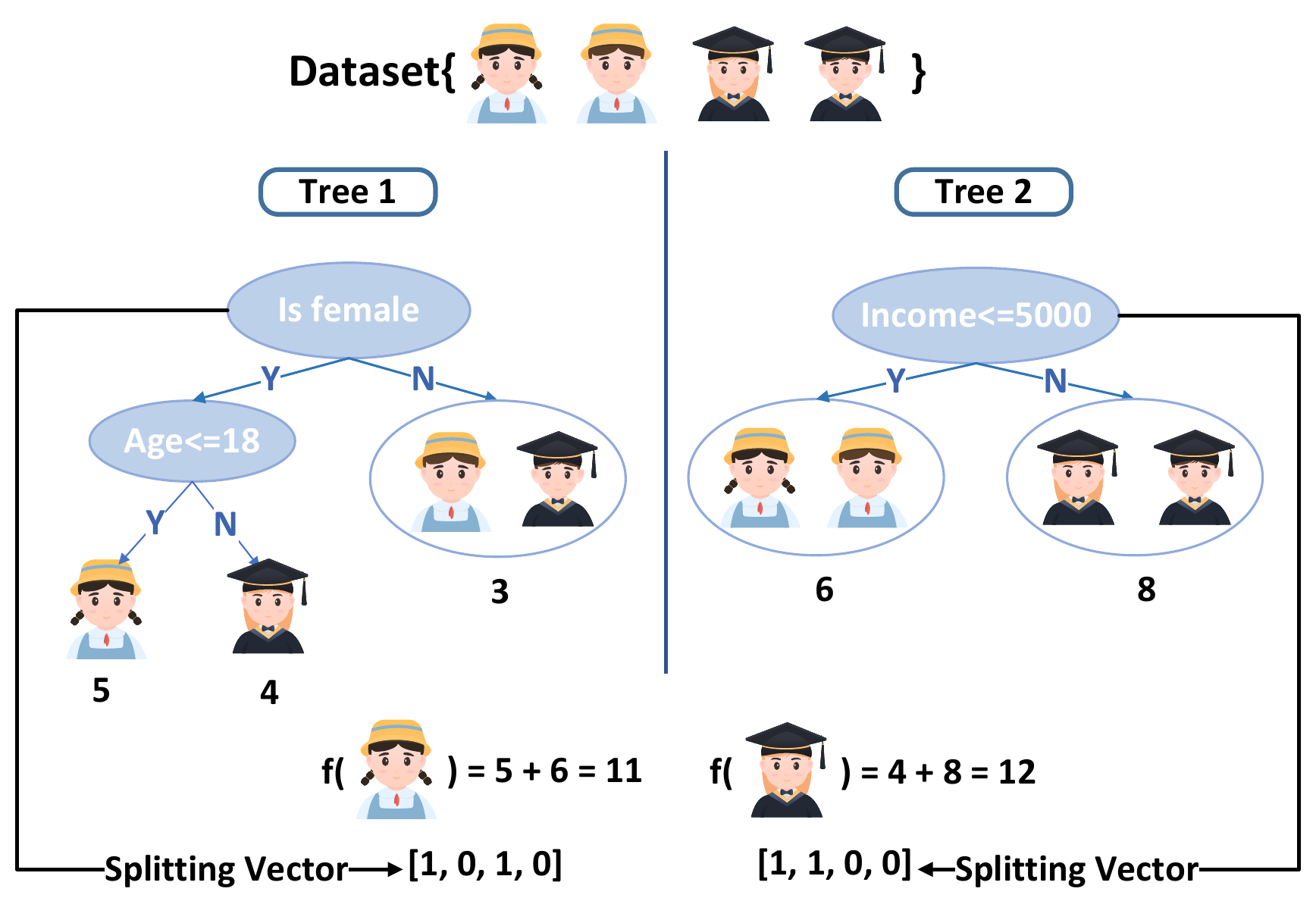}
\caption{An example of XGBoost and splitting vector. The final prediction for instance is the sum of the predictions from each regression tree. At the first level of Tree1, we split the instances based on whether the individual is female or not, which generates the splitting vector $[1,0,1,0]$.}
\label{xgboost}
\end{figure}

\mypara{XGBoost}
XGBoost is a boosting-based machine learning algorithm that ensembles a set of decision trees.
\autoref{xgboost} shows an example of XGBoost.
Next, we take a closer look at the XGBoost training.
For a regression model $ \psi(\cdot)$ consisting of $T$ regression trees $f_t$, we choose a set $\mathcal F$ of admissible tree models as follows 
\begin{equation}
\label{regularizer}
    \hat{y}_i := \psi (x_i) = \sum_{t = 1}^T f_t(x_i), f_t \in \mathcal{F},x_i \in \mathcal{X}.
\end{equation}

With any differentiable convex loss function, $l: \mathbb{R}\times\mathbb{R} \rightarrow \mathbb{R}$, capturing the disagreement between the labels $y_i$'s and predicted outputs $\hat y_i$'s, the objective function ${L}_0(\psi)$ for the model training process can be defined as
\begin{equation} \label{eqn:costFuncL}
    {L}_0(\psi) = \sum_{i = 1}^nl(y_i,\hat{y}_i) + \sum_{t=1}^T \Omega(f_t),
\end{equation}
where $\Omega(f_t):= \gamma L^{(t)} + \frac{1}{2}\lambda\norm{w^{(t)}}^2$ is the regularization on the model complexity of  tree $f_t$ to avoid overfitting, $L^{(t)}$ is the number of leaves and $w^{(t)}$ is the leaf weight of regression tree $f_t$. 

When we train a regression model, an iterative optimization method can be used to minimize the objective \eqref{eqn:costFuncL}. 
At the $t^{th}$ iteration, the previously constructed trees $f_1,\dots,f_{t-1}$ remain unchanged, while a new tree $f_t$ is trained for the first time and added to the regression model.
Given the predicted output 
$\hat{y}^{(t-1)}_i = \sum_{k = 1}^{t-1}f_k(x_i)$ at the $(t-1)^{th}$ iteration, the objective at the $t^{th}$ iteration is formulated as
    \begin{equation} 
        L_0^{(t)} = \sum_{i = 1}^n l(y_i, \hat{y}^{(t-1)}_i+ f_t(x_i))  + \Omega(f_t).\\
    \end{equation}  

To make the training process computationally efficient, rather than the objective itself, we usually minimize a second-order approximation of the objective function as follows
\begin{small}
\begin{equation} 
\label{eqn:optimizeFunc}
    {{L}}^{(t)}
     = \sum_{i = 1}^n \left(l(y_i, \hat{y}^{(t-1)}_i) + g_i^{(t)} f_t(x_i) + \frac{h_i^{(t)} f^{2}_t(x_i)}{2}\right) + \Omega(f_t),
\end{equation}
\end{small}where $g_i^{(t)} = \partial_{\hat{y}^{(t-1)}_i}l$ and $h_i^{(t)} = \partial^{2}_{\hat{y}^{(t-1)}_i}l$ are the first and  second derivative of the loss function at $\hat{y}^{(t-1)}_i$, respectively.

Let $\mathcal I_{l_j}$ denotes the index set of instances at the $j^{th}$ leaf, i.e., $I_{l_j} = \{i|x_i ~\text{belongs to} ~j^{th} ~\text{leaf of}~ f_t\}$. 
Chen~\cite{chen2016} has shown that the optimal weight of each leaf $j$ is computed by
\begin{equation} \label{eqn:optimalWeight}
    w^{(t),*}_j = - \frac{\sum_{i \in I_{l_j}} g^{(t)}_i}{\sum_{i \in I_{l_j}} h_i + \lambda}.
\end{equation}

The following metric is often used to evaluate a \textit{splitting candidate} at each node $j$ (associated with instance set $\mathcal I_j$) ~\cite{chen2016}:
\begin{footnotesize}\begin{equation} \label{eqn: LoptimalSplit}
\begin{aligned}
& {L}_{split}(\mathcal I_{L,j}, \mathcal I_{R,j}) = - \gamma + \\ 
&\, \frac{1}{2}\left( \frac{(\sum_{i \in I_{L,j}} g_i^{(t)})^2}{\sum_{i \in I_{L,j}} h_i^{(t)} + \lambda} + \frac{(\sum_{i \in I_{R,j}} g_i^{(t)})^2}{\sum_{i \in I_{R,j}} h_i^{(t)} + \lambda} - \frac{(\sum_{i \in I_j} g_i^{(t)})^2}{\sum_{i \in I_j} h_i^{(t)} + \lambda} \right), 
\end{aligned}
\end{equation}
\end{footnotesize}where $\mathcal I_{L,j}$ and $ \mathcal I_{R,j}$ are associated with the left and right child nodes, forming the dichotomy of the proposed splitting in question.
We refer the readers to Appendix \ref{xgboostappendix} for more details about XGBoost and the regression tree.

\mypara{Splitting Vector}
Next, we will illustrate XGBoost in another way in terms of the so-called splitting vector~\cite{le2021fedxgboost}. 
Let $d= [x_{1\boldsymbol{\cdot}},\ldots,x_{n\boldsymbol{\cdot}}]^{\top} \in \mathbb{R}^n $ be the aggregation of a particular entry of the feature vectors for all instances and $s_i \in \mathbb{R}$ is the $i^{th}$ splitting candidate for $d$ to split the instance set into two separate sets at a node.   
The splitting operation 
$\text{Split}(d,s_i): \mathbb{R}^n \times \mathbb{R} \longrightarrow \{0,1\}^{n}$ performs the comparison of each element in $d$ with the splitting candidate $s_i$ and outputs the $i^{th}$ splitting vector $ {m}_i \in \{0,1\}^{n}$ as
\begin{center}
$ {m}_i := \text{Split}(d, s_i) :=  [m_{1i},\ldots,m_{ni}]^{\top}$,
\end{center}
where ${m}_{ji} = \twopartdef{1,}{x_{j\boldsymbol{\cdot}} \leq s_i}{0,}{x_{j\boldsymbol{\cdot}} > s_i}$.

There corresponds a splitting operation to the splitting vector ${m}_i$ for a splitting candidate $s_i$, which labels any instances as ``1" if they are assigned to the left node and ``0"  to the right node. The following example depicts the functionality of the splitting operator and the splitting vector.

    {\example{} \label{example1} Consider a bank that holds data on many dimensions of its users, such as ages, genders, deposits, credit information, etc. 
    These data can reflect the economic situation of users, which is helpful for predicting the purchasing powers of users. 
    In \autoref{xgboost}, the bank builds the XGBoost model using ages, genders, and income information.
    Suppose we were analyzing the effect of a user's age. 
    Let the feature vector be $d = [24, 25, 20, 22, 15, 17, 18, 16]^{\top}$ representing the age data of a group of people. 
    A splitting candidate $s$ is chosen as Age = 18. 
    The result of the splitting operator applied to the feature vector $d$ and the candidate $s$ is denoted by the following splitting vector
    $$ m = \text{Split}(d, s) =  [
    0,0,0,0,1,1,1,1
   ]^{\top}.$$
    ``1" means that person is no older than 18, and ``0" means older than 18. 
    The splitting operation by $s$ (age = 18) is also depicted in \autoref{xgboost}. 
}

With the formulation of the splitting vector, the aggregated gradient and Hessian in each node can be computed by multiplying the splitting vector and gradient and Hessian vector. 
Let the gradient vector and Hessian vector of the $n$ instances respectively be denoted as
$$
g = [g_{1},\ldots,g_{n}]^{\top}, h = [h_{1},\ldots,h_{n}]^{\top} \in \mathbb{R}^n.
$$
To evaluate a splitting candidate $s$, we need to know the aggregated gradients and Hessians for the split instances at the left and right nodes, termed $g_s^L$, $h_s^L$, and $g_s^R$, $h_s^R$, respectively, which are computed from $m, g, h$ as follows:
\begin{equation}\label{eqn:computeSumGH}
g_s^L = m^{\top}g, ~h_s^L  = m^{\top}h, ~g_s^R = G - g_s^L,~h_s^R = H - h_s^L,
\end{equation}
where $ G := \sum_{i=1}^n g_i$  and $ H := \sum_{i=1}^n h_i$.
Then we can obtain the score ${L}^{s}_{split}$ of the splitting rule candidate as
\begin{equation} \label{eqn:Lsplit_short}
 {L}^{s}_{split} = - \gamma + \frac{1}{2}\left( \frac{(g^L_{s})^2}{(h^L_{s}) + \lambda} + \frac{(g^R_{s})^2}{(h^R_{s}) + \lambda} - \frac{G^2}{H + \lambda} \right).
\end{equation}
The essential of the training process for splitting at a node is to find the best splitting candidate $s$ with the highest score in \eqref{eqn:Lsplit_short}.
Splitting is repeated from the newly constructed nodes until it reaches the maximum depth of the tree.  
The so-constructed nodes together form a whole tree structure.

\begin{table}[!tbp]
	\centering 
	\caption{Summary of notations.} 
	\label{notationstable} 
    \setlength{\tabcolsep}{1mm}
	\begin{tabular}{c|l}  
        \toprule
		\textbf{Notations} & \textbf{Descriptions} \\ 
        \midrule
		$n$ & Number of the instances in a dataset \\
        $m_i$ & $i^{th}$ splitting vector \\
        $M$ & Categorical matrix \\
        $g,h$ & Gradient and Hessian vector \\
        $g_i,h_i$ & Gradient and Hessian of $i^{th}$ instance\\
        $W$ & Number of noise vectors for each $m_i$\\     
        $b_{ij}$ & $j^{th}$ noise vector for $m_i$\\
        $B_{i}$ & $i^{th}$ noise matrix for $m_i$\\ 
        ${\left\langle g \right\rangle}_i,{\left\langle h \right\rangle}_i$  & Noised gradient and Hessian vector for $m_i$ \\
        $\varepsilon_{\rm AP},\varepsilon_{\rm PP}$  & Privacy budgets for active party and passive party \\
        \bottomrule
	\end{tabular}
\end{table}

\subsection{Differential Privacy (DP) and Local Differential Privacy (LDP)}

\mypara{Differential Privacy}
\textit{Differential privacy} (DP) is a formal definition of privacy that guarantees the output of a data analysis does not depend significantly on a single individual’s data item. 
DP has been widely used in various fields, including control systems~\cite{le2013ddpfilter, huang2024differential}, machine learning~\cite{abadi2016deep}, data synthesis~\cite{ZWLHBHCZ21, WZWHBCZ23, YZDCCS23}, etc.

\begin{definition}[Differential Privacy~\cite{dwork2014algorithmic}] A randomized mechanism $\mathcal{M}$ satisfies $(\varepsilon, \delta)$-differential privacy $((\varepsilon, \delta)$-DP), where $\varepsilon \geq 0$ and $0 \leq \delta \leq 1$, if for any two datasets inputs $d$ and $d^{\prime}$ that differ in a single record and any measurable set $\mathcal O$ of $\mathcal{M}$'s outputs,
    \[{\bf Pr}(\mathcal{M}(d) \in\mathcal O) \leq \exp(\varepsilon){\bf Pr}(\mathcal{M}(d^{\prime}) \in\mathcal O) +\delta. \] 
\end{definition}
    
Here $\varepsilon$ is the \textit{privacy budget}. 
To achieve $(\varepsilon, \delta)$-differential privacy, the Gaussian and Laplace mechanisms are usually adopted by adding noise calibrated to the sensitivity of a function~\cite{dwork2006our}.
The Laplace mechanism can achieve strict $\varepsilon$-DP, while the Gaussian mechanism can only achieve $(\varepsilon, \delta)$-DP.
When $\delta = 0$, the mechanism $\mathcal{M}$ satisfies pure (strict) differential privacy (pure DP), namely, $\varepsilon$-DP. 
When $\delta > 0$, the mechanism satisfies approximate (relaxed) differential privacy (approximate DP), namely, $(\varepsilon, \delta)$-DP~\cite{dwork2006our}. 

The following results are fundamental to the privacy analysis when multiple randomized functions are applied to a dataset.

\begin{lemma}[Sequential Composition~\cite{dwork2014algorithmic}]
Let $\mathcal M=$ $\left\{\mathcal M_1, \ldots, \mathcal M_k\right\}$ be a series of randomized mechanisms performed sequentially on a dataset. If $\mathcal M_i$ provides $(\varepsilon_i,\delta_i)$-DP, then $\mathcal M$ provides $(\varepsilon, \delta)$-DP with $\varepsilon:=\sum_{i=1}^k \varepsilon_i$ and $\delta:=\sum_{i=1}^k \delta_i$.
\label{DPsequential}
\end{lemma}

\begin{lemma}[Parallel Composition~\cite{dwork2014algorithmic}]
Let $\mathcal M=$ $\left\{\mathcal M_1, \ldots, \mathcal M_k\right\}$ be a series of mechanisms performed separately on disjoint subsets of the entire dataset. If $\mathcal M_i$ provides $(\varepsilon_i,\delta_i)$-DP, then $\mathcal M$ provides $(\varepsilon, \delta)$-DP with $\varepsilon:=\max \{\varepsilon_1, \ldots, \varepsilon_k\}, \delta:=\max \{\delta_1, \ldots, \delta_k\}$.
\label{DPparallel}
\end{lemma}

For a sequential of $k$ mechanisms $\mathcal{M}_1, \ldots, \mathcal{M}_k$ satisfying $\left(\varepsilon_i, \delta_i\right)$-DP respectively, the basic sequential composition result~\cite{dwork2014algorithmic} shows that the privacy composes linearly, i.e., the sequential composition satisfies $\left(\sum_i^k \varepsilon_i, \sum_i^k \delta_i\right)$-DP. 
When $\varepsilon_i = \varepsilon$ and $\delta_i=\delta$, the advanced composition bound from~\cite{dwork2010boosting} states that the composition satisfies $\left(\varepsilon \sqrt{2 k \log \left(1 / \delta^{\prime}\right)}+k \varepsilon\left(e^{\varepsilon}-1\right)\right.$, $\left.k \delta+\delta^{\prime}\right)$-DP, where $\delta^{\prime} > 0$.

\mypara{Local Differential Privacy}
A stronger notion of DP is \textit{local differential privacy} (LDP), which perturbs the individual's data locally to protect private information~\cite{DHZFCZG23,DZBLJCC21,ZWLHC18,WCZSCLLJ21}. 
LDP is a model of differential privacy with the added restriction that even if an adversary has access to the personal responses of an individual in the database, that adversary is still unable to learn too much about the user's personal data.
This is contrasted with DP, which incorporates a central aggregator with access to the raw data. 

\begin{definition}[Local Differential Privacy~\cite{erlingsson2014rappor}] A randomized mechanism $\mathcal{M}$ satisfies $(\varepsilon, \delta)$-local differential privacy $((\varepsilon, \delta)$-LDP), where $\varepsilon \geq 0$ and $0 \leq \delta \leq 1$, if for any two of input $d, d^{\prime}$, and any measurable set $\mathcal O$ of $\mathcal{M}$'s outputs, 
\[{\bf Pr}(\mathcal{M}(d) \in \mathcal O) \leq \exp(\varepsilon){\bf Pr}(\mathcal{M}(d') \in \mathcal O) +\delta.\]
\end{definition}
 
\subsection{Notations}
Frequently used notations are summarised in \autoref{notationstable}.

\section{Problem Statement and Existing Solutions} \label{problem state}

\begin{figure} [!tbp]
\centering  
\includegraphics[width=0.45\textwidth]{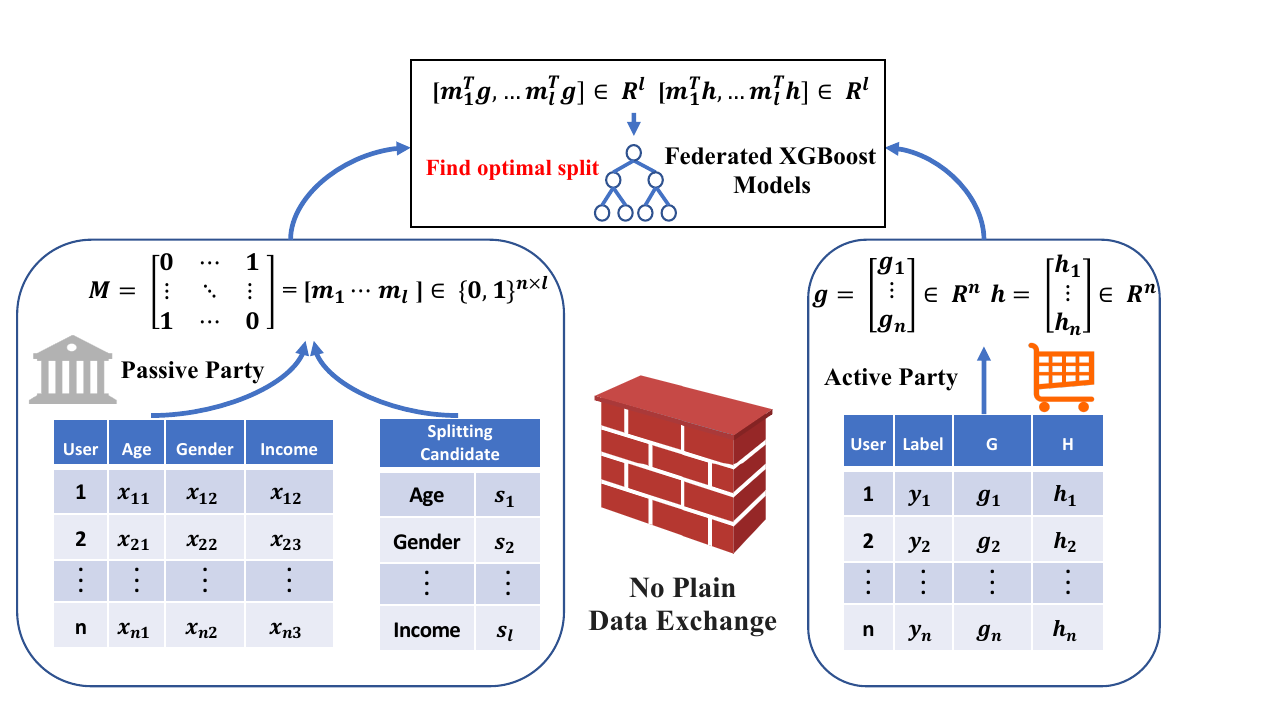}
\centering  
\caption{Federated XGBoost problem setting.}
\label{fig:SMM}
\end{figure}

\begin{figure*} [!tbp]
\centering  
\includegraphics[width=0.9\textwidth]{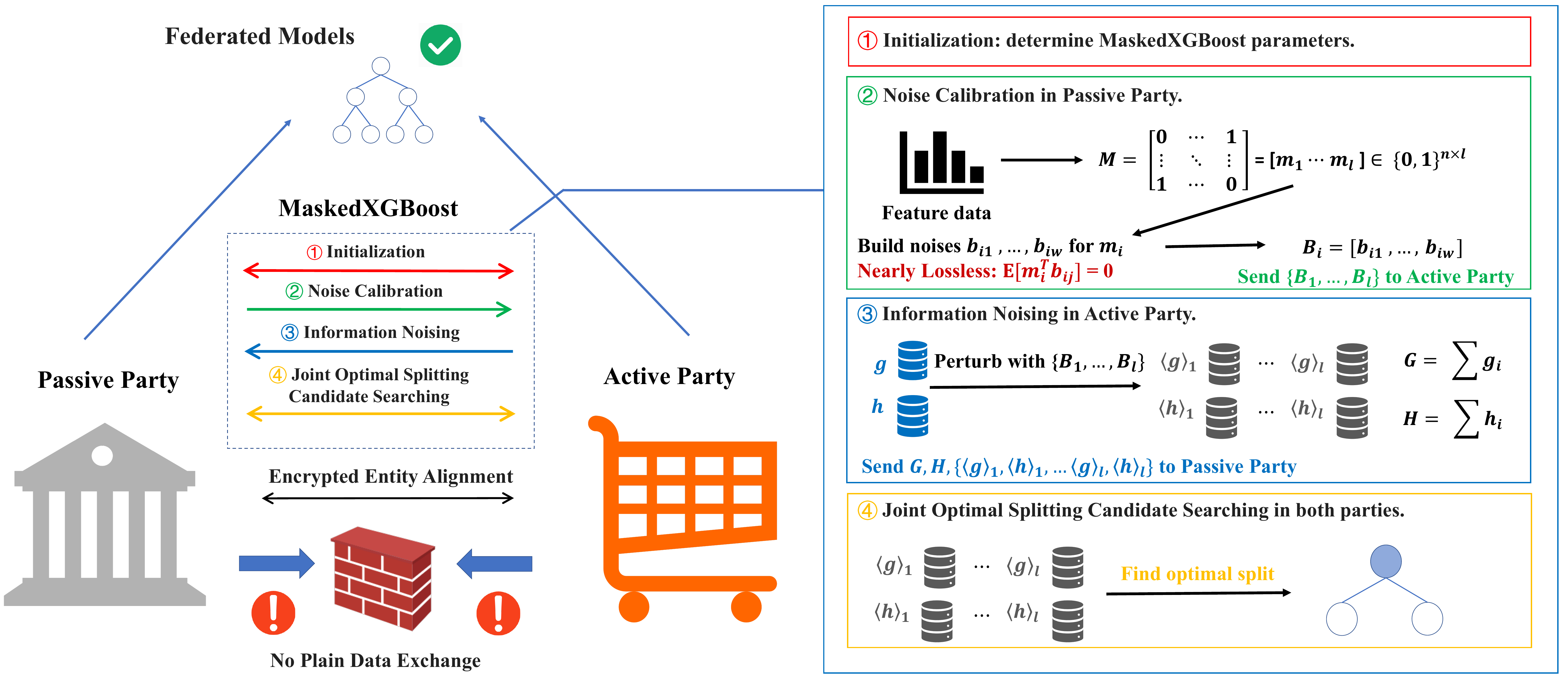}
\caption{\method overview. 
\method is composed of four steps: initialization, noise calibration, information noising, and joint optimal splitting candidate searching.
AP and PP first determine the necessary algorithm parameters in the initialization step.
In the noise calibration step, PP builds well-calibrated noise according to the categorical matrix which ensures the consistently better property of \method. 
In the information noising step, AP uses the noises PP sent to perturb the sensitive information $g$ and $h$, then AP sends the information back to PP. 
Finally, AP and PP jointly find the optimal splitting candidate. 
Then, splitting continues from the newly constructed nodes until it reaches the maximum depth of the tree. 
AP and PP stand for active party and passive party respectively.
}
\label{maskedxgboost_figure}
\end{figure*}

\subsection{Problem Statement}
\mypara{Federated XGBoost}
In this paper, we focus on VFL-setting XGBoost.
We call a party that holds both feature data and the class label \textit{active party} (AP) and call a party that merely holds the feature data \textit{passive party} (PP).
We consider that AP and PP want to train an XGBoost model jointly.
The essential part of the training state is computing the splitting score \eqref{eqn:Lsplit_short} and then finding the optimal splitting candidate. 
Assume that at each node, PP has $l$ splitting candidates, through each splitting vector $m_i \in \mathbb R^n$, the categorical matrix $M$ is defined as $M = [m_1,\ldots, m_l] \in \mathbb R^{n \times l}$. 
According to \eqref{eqn:computeSumGH}, AP has $g,h$, while PP owns $M$ and they desire to compute $M^{\top}g, M^{\top}h$ securely. Gradient and Hessian vectors in plain text cannot be sent directly, otherwise, users' label information from AP will be seriously leaked~\cite{cheng2021secureboost}.  
\autoref{fig:SMM} illustrates the problem setting. 
With $M^{\top}g, M^{\top}h$ known, the optimal splitting candidate can be found and the federated XGBoost model can be built step by step. 

{\remark Multi-party private set intersection~\cite{kolesnikov2017practical} is employed to determine the intersection of common databases among participants securely, which is orthogonal to our paper. 
The primary challenge in our setting lies in how to jointly conduct the training process among participants with the aligned database.
}

\mypara{Threat Model and Design Goals} 
In common with many other works in the federated setting~\cite{cheng2021secureboost,maddock2022hltreewithdp, li2021label, zhang2024s2nerf}, we assume an honest-but-curious model, where the clients do not trust others with their raw data.
Given the above setting, we aim to propose a federated XGBoost protocol with the following design goals:

\begin{itemize}
    \item \mypara{G1:~Utility}
    Its utility should be close to centralized learning, which is to pool all data into a centralized server.
    \item \mypara{G2:~Privacy}
    Its privacy should be close to local training, i.e., each participant trains with its local data only. 
    To achieve this, all data being transmitted should be protected either by cryptographic technology or differential privacy.
    \item \mypara{G3:~Efficiency}
    Its efficiency should be close to traditional distributed ML~\cite{chen2016,fu2019experimental, ke2017lightgbm}, i.e., the number of cryptographic operations should be minimized.
\end{itemize}

\subsection{Existing Solutions}

\mypara{HE Methods}
Current approaches that rely on Homomorphic Encryption (HE) are proficient in safeguarding privacy but often lack efficiency~\cite{cheng2021secureboost, chen2021secureboost+}.
At the same time, these methods lack the protection of PP's private feature information.
These methods can satisfy \textbf{G1}~(Utility) and \textbf{G2}~(Privacy) of only one party AP, while they cannot satisfy \textbf{G2}~(Privacy) of the other party PP and \textbf{G3}~(Efficiency).

\mypara{DP Methods}
The methods~\cite{tian2023sf,le2021fedxgboost} using DP have a great improvement in efficiency, but the utility is greatly disturbed.
The work~\cite{wang2022feverless} considers the VFL settings for decentralized labels, utilizing secure aggregation and differential privacy to protect the privacy.
Since only AP owns the label in our two-party VFL setting, this method is equivalent to adding noise directly to the gradient information.
These methods can satisfy \textbf{G2}~(Privacy) of both parties and \textbf{G3}~(Efficiency), with serious sacrifice of \textbf{G1}~(Utility).

\mypara{Other Methods}
In the splitting vector view of XGBoost, Le et al.~\cite{le2021fedxgboost} proposed FedXGBoost-SMM.
FedXGBoost-SMM turns the federated XGBoost into a secure matrix multiplication problem, and its key idea is to build kernel vectors as noise vectors. 
FedXGBoost-SMM can not provide a differential privacy guarantee and finding kernel vectors for a large categorical matrix is still time-consuming.
FedXGBoost-SMM can satisfy \textbf{G1}~(Utility) with the sacrifice of \textbf{G2}~(Privacy) and \textbf{G3}~(Efficiency).

\section{Our Proposal: \method} 
\label{maskedxgb}


\subsection{Overview}

The general idea of \method is that PP generates well-calibrated noises according to the categorical matrix $M$, and AP uses these noises to perturb sensitive vectors $g,h$, and sends the masked vectors back to PP. 
Then PP multiplies the categorical matrix and the masked vectors to find an optimal splitting candidate. 
The whole process of the protocol includes four steps: Initialization, noise calibration, information noising, and joint optimal splitting candidate searching.
\autoref{maskedxgboost_figure} illustrates the main steps of \method.

\begin{algorithm}[!tbp]
	\caption{Noise Calibration}
	\label{alg:Secure-Noise}
	\textbf{Input:} Private categorical matrix $M = [m_1,\ldots,m_{l}] \in \mathbb{R}^{n \times l}$, $ m_i \in \{0,1\}^{n}$, \method parameters $\sigma_1, \sigma_2 , W > 0$ \\
    \textbf{Output:} Noise matrices set $\{B_1, \ldots B_l\}$ \\
    \textbf{Procedures:}
    \begin{algorithmic}[1]
        \FOR{$i = 1$ to $l$} 
            \FOR{$j = 1$ to $W$} 
                \STATE $u_{ij}, v_{ij}, r_{ij} \leftarrow \{0\}^n$
                \STATE Generate $p_1,\ldots,p_{n_{A,i}}\sim \mathcal{N}\left(0, \sigma_1^2 \right)$, $q_1,\ldots,q_{n_{I,i}}\sim \mathcal{N}\left(0, 2\sigma_1^2 \right)$, $r_{ij}\sim \mathcal{N}\left(0, \sigma_2^2 I \right)$
                \FOR{$k = 1$ to $n_{A,i}$} 
                    \IF{$k = 1$}
                        \STATE $[u_{ij}]_{\mathcal{A}(m_i)_k} = p_1 - p_{n_{A,i}}$
                    \ELSE
                        \STATE $[u_{ij}]_{\mathcal{A}(m_i)_k} = p_k - p_{k-1}$
                    \ENDIF                    
                \ENDFOR
                \FOR{$k = 1$ to $n_{I,i}$} 
                    \STATE $[v_{ij}]_{\mathcal{I}(m_i)_k} = q_k$            
                \ENDFOR                
                \STATE $b_{ij} \leftarrow  u_{ij} + v_{ij} + r_{ij}$
            \ENDFOR
            \STATE $B_i \leftarrow [b_{i1},\ldots,b_{iW}]$
        \ENDFOR
    \end{algorithmic}
\textbf{Return:} Noise matrices set $\{B_1, \ldots, B_l\}$
\end{algorithm}

\mypara{(S1) Initialization} Before the protocol begins in earnest, the two parties negotiate a level of differential privacy guarantee. Then according to this privacy setting, AP and PP can get the suitable parameters $\sigma_1,\sigma_2, C, W$ which will be applied to the concrete procedures in the following steps, with the guidance of \autoref{theorem:utility}, \autoref{theorem:APLDP}, and \autoref{theorem:PPDP}, which will be introduced in more detail in \autoref{theoreticalanalysis}.
Before the whole training, AP randomly initializes the predicted labels and computes the initial gradients and Hessians~\cite{cheng2021secureboost}.

\mypara{(S2) Noise Calibration in PP} In this step, PP builds noises calibrated to the categorical matrix $M$ and then sends the noises to AP. 
This step will be introduced in detail in \autoref{subsec:noise_calibration}.

\mypara{(S3) Information Noising in AP}
After receiving the noises PP sent, AP uses these noises to perturb the sensitive information and then sends them to PP for the next step.
This step will be introduced in detail in \autoref{subsec:information_perturbation}.

\mypara{(S4) Joint Optimal Splitting Candidate Searching}
With the information AP sent, PP can evaluate splitting candidates and then find the optimal splitting candidate according to \eqref{eqn:Lsplit_short}. 
PP is requested to reveal the corresponding splitting score and operation. 
AP finds the best splitting candidate over AP and PP.
Then AP constructs new tree nodes and repeats the process with the new set of users.

The \method algorithm repeats steps \textbf{S2-S4} until a tree reaches the maximum depth, and builds new trees till the maximum training rounds.
The predicted labels are updated based on the evolving tree structure, and gradients and Hessians are updated accordingly.

\subsection{Noise Calibration} 
\label{subsec:noise_calibration}

In this step, PP builds noise vectors according to the categorical matrix $M$. 
We combine the Gaussian mechanism to construct orthogonal vectors of each splitting vector $m_i$ and then take these orthogonal vectors as noise vectors.

\mypara{Active Set and Inactive Set} 
To facilitate the presentation, we give the definition of the active set and inactive set. 
The \textit{active set} refers to the index set of a splitting vector that has a value of ``1", while the \textit{inactive set} refers to the index set that has a value of ``0".
For each splitting vector $m_i \in \{0,1\}^{n}$, we define the active set $\mathcal{A}(m_i)$ and the inactive set $\mathcal{I}(m_i)$ as follows:
$$
\begin{aligned}
\mathcal{A}(m_i) &:=\left\{j \in \mathcal{I} \mid m_{ij} = 1\right\}. \\
\mathcal{I}(m_i) &:= \mathcal{I} \backslash \mathcal{A}(m_i)=\left\{j \in \mathcal{I} \mid m_{ij} = 0\right\}.
\end{aligned}
$$
We denote $n_{A,i} = |\mathcal{A}(m_i)|$ and $n_{I,i} = |\mathcal{I}(m_i)|$. The active set includes all the nonzero entries of a splitting vector, while the inactive set encapsulates the zero.
For noise calibration, We will combine differently structured noises tailored to the active and inactive sets, namely the active and inactive noises.

\mypara{Noise Calibration}
\autoref{alg:Secure-Noise} describes the procedure of building noises. 
For each $m_i$, we build a noise matrix $B_i = [b_{i1},\ldots,b_{iW}]$, where $b_{ij} \in \mathbb R^n$ consists of three parts, \textit{active noise $u_{ij}$}, \textit{inactive noise $v_{ij}$}, and \textit{disturbing noise $r_{ij}$}, i.e. $b_{ij} = u_{ij} + v_{ij} + r_{ij}$. 
The active noise perturbs the bits of the active set in the gradient and Hessian vector. 
It is spatially colored as it is in the null space of $m_i$.
The inactive noise perturbs the bits of the inactive set in the gradient and Hessian vector freely. The disturbing noise perturbs all the bits of the gradient and Hessian vector together.
It is critical for the privacy protection of PP's categorical matrix.  
Next, we introduce how the three kinds of noises are constructed. 

\mypara{Active Noise $u_{ij}$}
Generate i.i.d. random numbers $p_1,\ldots,p_{n_{A,i}}$ following Gaussian distribution $\mathcal{N}(0, \sigma_1^2 )$.
We use $\mathcal{A}(m_i)_k$ to denote the $k^{th}$ element in the set $\mathcal{A}(m_i)$. 
Firstly, we initialize $u_{ij} \in\mathbb R^n$ as $0$, and 
then assign values to $u_{ij}$ iteratively:
\begin{center}
$[u_{ij}]_{\mathcal{A}(m_i)_k} = \twopartdef{p_1 - p_{n_{A,i}},}{k = 1}{p_k - p_{k-1},}{k = 2,\ldots, n_{A,i}}.$ 
\end{center}
It can be checked that $
m_i^{\top}u_{ij} = 0$ since the elements in the active noise $u_{ij}$ cancel each other out. 
The active noise does not affect the calculation result of $m_i^\top g$ and $m_i^\top h$ if used for information noising, indicating that this part of the noise is lossless.

\mypara{Inactive Noise $v_{ij}$}
Generate i.i.d. random numbers $p_1,\ldots,p_{n_{I,i}}$ following Gaussian distribution $\mathcal{N}(0, 2\sigma_1^2 )$.
We use $\mathcal{I}(m_i)_k$ to denote the $k^{th}$ element in the set $\mathcal{I}(m_i)$.
First we initialize the inactive noise as $v_{ij} =0$, then assign values to $v_{ij}$ iteratively:
\begin{center}
$[v_{ij}]_{\mathcal{I}(m_i)_k} = q_k, k = 1,\ldots, n_{I,i}$.
\end{center}
It is immediately known that $ m_i^{\top}v_{ij} = 0 $ because $v_{ij}$ only operates on the bits in the inactive set, which correspond to the bits in $m_i$ with a value of 0.
Similarly to the active noise, this implies that the inactive noise does not affect the calculation result of $m_i^\top g$ and $m_i^\top h$ too if used for information noising, revealing the lossless property of the inactive noise.

\mypara{Disturbing Noise $r_{ij}$}
The disturbing noise is generated much more easily. 
We generate $r_{ij}$ free from a multidimensional Gaussian distribution, $r_{ij}\sim \mathcal{N}(0, \sigma_2^2 I_{n} )$. 
It holds that
\begin{center}
$\mathbb{E}[m_i^\top r_{ij}] = 0$,  ${\rm Var}(m_i^\top r_{ij}) = n_{A,i}\sigma_2^2$.
\end{center}
This is because although $r_{ij}$ operates on all the bits, only the bits in the active set will perturb the outcomes effectively. 

Repeating the noise generation process above for $W$ times, PP builds $b_{i1}, \ldots, b_{iW}$ as a whole noise matrix $B_i$, i.e. $B_i = [b_{i1}, \ldots, b_{iW}] \in \mathbb{R}^{n \times W}$. 
Similarly, PP can generate $B_1, \ldots, B_l$ according to $l$ splitting vectors $m_1,\ldots,m_l$. 
Then PP sends the noise matrices set $\{B_1, \ldots, B_l\}$ to AP. 
The following example can visualize the noise generation procedure.

{\example{} \label{noiseexample}
We consider the setting in Example \autoref{example1}, assuming the splitting candidate $Age=18$ is the $i^{th}$ splitting candidate.
After analyzing the users' age information, PP builds the splitting vector $m_i = [0 ~0 ~0 ~0 ~1 ~1 ~1 ~1]^{T}$. 
The active set $\mathcal{A}(m_i)$ is $\left\{ 5,6,7,8 \right\}$, inactive set $\mathcal{I}(m_i)$ is $\left\{ 1,2,3,4 \right\}$. 
Generate i.i.d. random variables $p_1,\ldots,p_4$ according to $\mathcal N(0, \sigma_1^2)$ and 
$q_1,\ldots, q_4$ according to $\mathcal{N}(0, 2\sigma_1^2)$. 
Next we construct the first noise $b_{i1}$, which means constructing $u_{i1},v_{i1},r_{i1}$.

$$
\begin{gathered}
u_{i1} = \begin{bmatrix}~0~\\~p~\end{bmatrix}, v_{i1}  = \begin{bmatrix}~q~\\~0~\end{bmatrix}, r_{i1} \sim \mathcal{N}\left(0, \sigma_2^2 I\right) \\
\end{gathered}
$$
where 
$
p = [p_1-p_4, p_2-p_1, p_3-p_2, p_4-p_3]^{\top}$ and $
q = [q_1, q_2, q_3, q_4]^{\top}$. 
Overall, $b_{i1} = u_{i1} + v_{i1} + r_{i1}$. 
Now We use a numerical simulation to illustrate this procedure. Let $\sigma_1^2 = 1, \sigma_2^2 = 0.1$. Numerically generated noises are $p_1 = 0.9109, ~p_2 = -0.2397, ~p_3 = 0.1810, ~p_4 = 0.2442$, and 
$$
\begin{gathered}
u_{i1} = [0,0,0,0,0.6667, -1.1506, 0.4207, 0.0632]^{\top}, \\
v_{i1} = [-0.9613, 1.6737, 5.0767, -2.6467,0,0,0,0]^{\top}, \\
r_{i1} = [0.0138, -0.1907, -0.0365,    -0.0848, \\
-0.0765, -0.1128, 0.0078, 0.2107]^{\top}.
\end{gathered}
$$
The noise vector is 
$$
\begin{gathered}
b_{i1} = [-0.9475, 1.4830, 5.0402, -2.7315, \\
0.5902, -1.2634, 0.4285, 0.2739]^{\top}.
\end{gathered}
$$
Thus we can verify that
$$
\begin{gathered}
m_i^{\top}u_{i1} = 0, m_i^{\top}v_{i1} = 0, m_i^{\top}r_{i1} = 0.0292, m_i^{\top}b_{i1} = 0.0292
\end{gathered}
$$
We see that the active noise $u_{i1}$ and inactive noise $v_{i1}$ are lossless.
and that $b_{i1}$ follows a Gaussian distribution $\mathcal{N}(0, \Sigma )$, where $\Sigma =  diag(\Sigma_1, \Sigma_2)$ with $\Sigma_1 = 2.1I_{4}$ and 
$$ 
\Sigma_2 = 
\begin{bmatrix}
2.1 & -1 & 0 & -1 \\
-1 & 2.1 & -1 & 0\\
0 & -1 & 2.1 & -1 \\
-1 & 0 & -1 & 2.1
\end{bmatrix}
$$
}

\begin{algorithm}[!tbp]
	\caption{Information Noising}
	\label{alg:Secure-Response}
	\textbf{Input:} Noise matrices set $\{B_1,\ldots, B_l\}$, Private data $g,h \in \mathbb{R}^{n}$, \method parameter $C > 0$ \\
    \textbf{Output:} 
    $\left\{ \left\{  {\left\langle g\right\rangle}_i,{\left\langle h\right\rangle}_i \right\}_{i \in \left\{ 1, \ldots, l \right\}}, G, H\right\}$ \\
    \textbf{Procedures:} 
    \begin{algorithmic}[1]
        \FOR{$i = 1$ to $l$} 
            \STATE Generate $c_{i1},\ldots,c_{iW}$ and $d_{i1},\ldots,d_{iW}$ satisfy $\sum_{k = 1}^W c_{ik}^2 = \sum_{k = 1}^W d_{ik}^2 = C$
            \STATE ${\left\langle g \right\rangle}_i \leftarrow g + \sum_{k = 1}^W c_{ik}b_{ik}, {\left\langle h \right\rangle}_i \leftarrow h + \sum_{k = 1}^W d_{ik}b_{ik}$
        \ENDFOR
        \STATE $G \leftarrow \sum_{i = 1}^n g_{i}, H \leftarrow \sum_{i = 1}^n g_{i}$
    \end{algorithmic}
    
    \textbf{Return:} $\left\{ \left\{  {\left\langle g\right\rangle}_i,{\left\langle h\right\rangle}_i \right\}_{i \in \left\{ 1, \ldots, l \right\}}, G, H\right\}$
\end{algorithm}

\subsection{Information Noising} 
\label{subsec:information_perturbation}

\autoref{alg:Secure-Response} describes the process of noising sensitive information.
Concretely, AP has the sensitive information gradient and Hessian $g,h \in \mathbb{R}^{n}$ that needs to be perturbed. 
We use ${\left\langle g \right\rangle}_i$ and ${\left\langle h \right\rangle}_i$ to denote the noised gradient and Hessian for $i^{th}$ splitting candidate, respectively.
AP needs to generate ${\left\langle g \right\rangle}_i, {\left\langle h \right\rangle}_i \in \mathbb{R}^{n}$ for each splitting candidate $s_i$.

Next we introduce how ${\left\langle g \right\rangle}_i, {\left\langle h \right\rangle}_i$ are generated with $B_i = [b_{i1}, \ldots, b_{iW}] \in \mathbb{R}^{n \times W}$. 
AP generates $2W$ random numbers $c_{i1},\ldots,c_{iW}$ and $d_{i1},\ldots,d_{iW}$ with a constraint that 
$$
\sum_{k = 1}^W c_{ik}^2 = \sum_{k = 1}^W d_{ik}^2 = C.
$$
Then we perturb $g,h$ as 
\begin{equation}\label{def::purtured_g_h}
{\left\langle g \right\rangle}_i:= g + \sum_{k = 1}^{W} c_{ik}b_{ik}, {\left\langle h \right\rangle}_i := h + \sum_{k = 1}^{W} d_{ik}b_{ik}.
\end{equation}

So ${\left\langle g \right\rangle}_1,\ldots,{\left\langle g \right\rangle}_l$ and ${\left\langle h \right\rangle}_1,\ldots,{\left\langle h \right\rangle}_l$ can be generated iteratively. 
If PP wants to evaluate each splitting candidate, PP needs to know the sum of gradients and Hessians $G$ and $H$ in \eqref{eqn:Lsplit_short}.
So after noising, AP compute 
$$
G = \sum_{i=1}^n g_i, H = \sum_{i=1}^n g_i.
$$
Finally AP sends $\left\{ \left\{  {\left\langle g\right\rangle}_i,{\left\langle h\right\rangle}_i \right\}_{i \in \left\{ 1, \ldots, l \right\}}, G, H\right\}$ to PP. 

Intuitively, we can observe that increasing the value of $C$ leads to a higher perturbation being added to $g$ and $h$, resulting in stronger privacy protection.
We constrain the sum of $c_{ik}^2$ and $d_{ik}^2$ to control the level of perturbation of the information.
This will be analyzed in detail in \autoref{theoreticalanalysis}.

\begin{algorithm}[!tbp] 
	\caption{Split Finding}
    \label{alg:1}
	\textbf{Input:}
    \begin{itemize}
        \item Dataset $\mathcal{D}$, with $|\mathcal{D}| = n$ 
        \item \textbf{AP}: Private data $g, ~h \in \mathbb{R}^n$
        \item \textbf{PP}: Private categorical matrix $M = [m_1,\ldots,m_{l}] \in \{0,1\}^{n \times l}$
    \end{itemize}
    \textbf{Output of protocols:} 
    optimal splitting candidate $s^*$\\
    \textbf{Procedures:}
    \begin{algorithmic}[1]
		\STATE \textbf{PP}: $\{B_1,\ldots, B_l\}$ $\xleftarrow{\autoref{alg:Secure-Noise}}$Noise Calibration($M$)
		\STATE \textbf{PP}: Transmit $\{B_1,\ldots, B_l\}$ to AP.
		\STATE \textbf{AP}: $\left\{ \left\{  {\left\langle g\right\rangle}_i,{\left\langle h\right\rangle}_i \right\}_{i \in \left\{ 1, \ldots, l \right\}}, G, H\right\}$ $\xleftarrow{\autoref{alg:Secure-Response}}$ Information Noising ($\{B_1,\ldots, B_l \}$)
        \STATE \textbf{AP}: Transmit $\left\{ \left\{  {\left\langle g\right\rangle}_i,{\left\langle h\right\rangle}_i \right\}_{i \in \left\{ 1, \ldots, l \right\}}, G, H\right\}$ to PP.
		\STATE \textbf{PP}: \text{/* Compute the aggregated gradients and Hessians */} 
        \STATE $\{g^L_{s_i}, h^L_{s_i}\} \leftarrow \{m_i^{\top} {\left\langle g\right\rangle}_i, m_i^{\top}  {\left\langle h\right\rangle}_i \}$ 
		\STATE $L^* \leftarrow - \infty, ~s^* \leftarrow 0$
		\FOR{$i = 1$ to $l$} 
            \STATE $g^R_{s_i} \leftarrow G - g^L_{s_i}, ~h^L_{s_i} \leftarrow H - h^L_{s_i}$ \
            \STATE $L \leftarrow \frac{1}{2} \left( \frac{(g^L_{s_i})^2}{h^L_{s_i} + \lambda} + \frac{(g^R_{s_i})^2}{h^R_{s_i} + \lambda} - \frac{G^2}{H + \lambda}\right)$
            \IF{$L^* < L$}
                \STATE $L^* \leftarrow L, ~s^* \leftarrow s_i $
            \ENDIF
        \ENDFOR
        \STATE Transfer $L^*$ to AP, save $s^*$
    \\
    \STATE \textbf{AP:} Finds the optimal splitting candidate $s^*$ over AP and PP. 
    \end{algorithmic}
\end{algorithm}

\subsection{Putting Things Together}
With the information AP sent at the end of the information noising step,  PP evaluates splitting candidates and then finds the optimal splitting candidate according to \eqref{eqn:Lsplit_short}. 
\autoref{alg:1} describes the procedure of optimal split finding.
PP is requested to reveal the corresponding splitting score and splitting operation. 
AP finds the best splitting candidate according to the splitting score.
AP then constructs new nodes and repeats the process with the new set of users. 
In the next section, we will discuss how to set the \method's parameters $\sigma_1, \sigma_2, C, W$ considering both model utility and privacy guarantee.

{\remark Our algorithm can be adapted to the GBDT algorithm. GBDT is similar to XGBoost, but it only needs gradients to calculate the splitting score. We can protect the gradients by transmitting the noised gradients in \autoref{alg:Secure-Response} for the adaption of GBDT.}

\section{Theoretical Analysis} 
\label{theoreticalanalysis}

\subsection{Utility Analysis}
\label{subsec:utilitytheorem}
In this section, we show that \method is consistently better. 
Similar to the related work that adds noise to the adjusted score function \cite{le2021fedxgboost}, the performance of \method is related to the disturbance of the splitting scoring function 
\eqref{eqn:computeSumGH}. 
A higher level of disturbance in the scoring function increases the likelihood of failing to identify an optimal splitting candidate, which in turn leads to sub-optimal performance.

We show this by evaluating $i^{th}$ splitting candidate as an example. 
Because of the particularity of the added noise in \autoref{alg:Secure-Noise}, it can be checked that 
\begin{equation}\label{eqn::noised_product_outcomes}
m_i^{\top}{\langle g\rangle}_i = m_i^{\top}g + \sum_{k = 1}^{W} c_{ik}r_{ik},  m_i^{\top}{\langle h\rangle}_i = m_i^{\top}h + \sum_{k = 1}^{W} d_{ik}r_{ik},
\end{equation}
indicating that $m_i^{\top}{\langle g\rangle}_i$ and $m_i^{\top}{\left\langle h\right\rangle}_i$ are merely disturbed by  $r_{ik}$, making a better utility protocol possible.
Notice that the splitting score function becomes random due to the noised aggregated gradients and Hessians ${\langle g \rangle}_i, {\langle h \rangle}_i$. 
Define the newly noised splitting score function 
$\langle L_{split}^{s_i}\rangle$ as 
$$
\langle L_{split}^{s_i}\rangle = - \gamma + \frac{1}{2}\left( \frac{({\langle g^L_i \rangle})^2}{{\langle h^L_i \rangle} + \lambda} + \frac{({\langle g^R_i \rangle})^2}{{\langle h^R_i \rangle} + \lambda}  - \frac{G^2}{H + \lambda} \right).
$$
where ${\langle g^L_i \rangle} = m_i^{\top}{\langle g \rangle}_i, {\langle h^L_i \rangle} = m_i^{\top}{\langle h \rangle}_i$.
The utility of \method can be evaluated by the following concentration analysis.

{\theorem{} \label{theorem:utility}
	Denote $\kappa:=\sqrt{n_{A,i}C}\sigma_2$, there exists an upper bound $U(\alpha, \kappa)$ such that for any deviation $\alpha>0$:
	$$
	{\bf Pr}\left(\mid \langle L_{split}^{s_i}\rangle  - L_{split}^{s_i} \mid \geq \alpha\right)\leq U(\alpha, \kappa)
	$$
 and $\lim_{\kappa\rightarrow0}U(\alpha,\kappa)=0$. In particular, for    $\alpha\leq\operatorname{min}\left((\frac{(g_i^L)^2}{(h_i^L)+\lambda},\frac{(g_i^R)^2}{(h_i^R)+\lambda}\right)$, $U(\alpha,\kappa)$ is given by~\eqref{eqn:upper_bound}.\\
}
	\begin{figure*}[t]
	\begin{align}\label{eqn:upper_bound}
			U(\alpha,\kappa)= & \int_{0}^{\infty} \left(\Phi\left(\frac{-\sqrt{\beta_Lt}+\mu_X}{\kappa}\right)+\Phi\left(\frac{-\sqrt{\beta_Lt}-\mu_X}{\kappa}\right)+\Phi\left(\frac{\sqrt{\beta_L't}-\mu_X}{\kappa}\right)+\Phi\left(\frac{\sqrt{\beta_L't}+\mu_X}{\kappa}\right)\right)\frac{1}{\kappa}\phi\left(\frac{t-\mu_Y}{\kappa}\right){\rm d}t\notag\\
			&+\int_{0}^{\infty} \left(\Phi\left(\frac{-\sqrt{\beta_Rt}+\mu_Z}{\kappa}\right)+\Phi\left(\frac{-\sqrt{\beta_Rt}-\mu_Z}{\kappa}\right)+\Phi\left(\frac{\sqrt{\beta_R't}-\mu_Z}{\kappa}\right)+\Phi\left(\frac{\sqrt{\beta_R't}+\mu_Z}{\kappa}\right)\right)\frac{1}{\kappa}\phi\left(\frac{t-\mu_W}{\kappa}\right){\rm d}t\notag\\
			&+\Phi\left(-\mu_Y/\kappa\right)+\Phi\left(-\mu_W/\kappa\right)\notag\\
		&\beta_L:=\frac{(g_i^L)^2}{(h_i^L)+\lambda}+\alpha,\quad \beta_L':=\frac{(g_i^L)^2}{(h_i^L)+\lambda}-\alpha,\quad\beta_R:=\frac{(g_i^R)^2}{(h_i^R)+\lambda}+\alpha,\quad\beta_R':=\frac{(g_i^R)^2}{(h_i^R)+\lambda}-\alpha\notag \\
			&\mu_X:=(g_i^L),\quad \mu_Y:=(h_i^L)+\lambda,\quad \mu_Z:=(g_i^R),\quad \mu_W:=(h_i^R)+ \lambda
	\end{align}
  with standard normal density $\phi$ and cumulative distribution function $\Phi$.
  
	\hrulefill
\end{figure*}

The proof can be found in Appendix \ref{losslessappendix}.
From~\eqref{eqn::noised_product_outcomes}, we see that \method's performance is related to $C$ and $\sigma_2$, and it has nothing to do with $\sigma_1$. Theorem~\ref{theorem:utility} supports this hypothesis well. 
We claim that \method achieves better utility in the sense that we can choose $\sigma_2$ and $C$ to make the error bound $U(\alpha,\kappa)$ for estimating the splitting score sufficiently small.
In \autoref{ablation}, we also test the effect of $\sigma_1$ and $\sigma_2$ on \method's performance numerically.
Next, we introduce how to set reasonable parameters $C$, $\sigma_1$, and $\sigma_2$ so that \method satisfies the differential privacy guarantee for both parties of federated scenarios.

\subsection{Privacy Analysis: Differential Privacy at Active Party Side} \label{APprivacy}

According to our protocol described in \autoref{maskedxgb}, AP needs to send the masked ${\langle g \rangle}_i, {\langle h\rangle}_i$ to PP, which can be used to discover the class label information~\cite{cheng2021secureboost}. 
So we need to analyze the privacy preservation of AP. 

For each splitting vector $m_i$, AP uses the noises $b_{i1},\ldots, b_{iW}$ received from PP to perturb $g,h$ according to \autoref{alg:Secure-Response}. 
This is equivalent to adding a noise vector $\eta_i$ directly to $g,h$, $\eta_i \sim \mathcal{N}\left(0, \Sigma \right)$, where the submatrix with rows and columns indexed by $\mathcal{A}(m_i)$ is given as
\begin{align}\label{eqn:signma_active_set}
&\Sigma_{\mathcal{A}(m_i)} = \\
&\begin{bmatrix}
C(2\sigma_1^2 + \sigma_2^2) & -C\sigma_1^2 & 0& \cdots& 0 & -C\sigma_1^2 \\
-C\sigma_1^2 &\ddots & \ddots & \ddots & & 0\\
0 & & \ddots & \ddots & \ddots&  \vdots\\
\vdots& & &\ddots & \ddots & 0\\
0 & & & & \ddots& -C\sigma_1^2 \\
-C\sigma_1^2 & 0 &\cdots & 0& -C\sigma_1^2 & C(2\sigma_1^2 + \sigma_2^2)
\end{bmatrix}\notag
\end{align}
and the submatrix of all the other entries is given as
\begin{equation}\label{eqn:signma_inactive_set}
\Sigma_{\mathcal{I}(m_i)} = C(2\sigma_1^2+\sigma_2^2)I.
\end{equation}
Notice that 
$\Sigma_{\mathcal{A}(m_i)} $ is a circulant matrix~\cite{davis1979circulant}. 
We employ LDP to evaluate the privacy preservation of our protocol.

To perform the LDP analysis, it is essential to bound the sensitivity of the query function. In our context, this requires establishing upper bounds for both the gradients and Hessians.
For binary classification problems, the gradient and Hessian are bounded, i.e.,
there exists a $\mu>0$ such that $|g_i|,|h_i|<\mu/2$, and $\mu$ can be conveniently derived.
Specifically, the loss function is of the form $l(y_i,\hat{y_i}) = y_i \log(\hat{y_i}) + (1-y_i) \log(1-\hat{y_i})$ (i.e., binary cross-entropy).
We have gradients $g_i \in \left[-\frac{1}{2},-\frac{1}{e+1} \right] \cup \left[\frac{1}{2},\frac{e}{e+1} \right]$ and Hessians $h_i \in \left[ 0, \frac{1}{4}\right]$; thus $\mu = \frac{2e}{e+1}$.
For other problems (e.g., regression problems), the loss functions may have unbounded gradient and Hessian. 
We may use gradient clipping (similar to DP-SGD~\cite{abadi2016deep}) to obtain bounded gradients and Hessians and then derive $\mu$.

{\theorem{} 
Recall that $n_{A,i}$ and $n_{I,i}$ are the sizes of the active set and the inactive set of the splitting vector $m_i$ respectively, \method is $(\varepsilon_{\rm AP}, \delta_{\rm AP})$-local differentially private, where $\varepsilon_{\rm AP} > 0, 0<\delta_{\rm AP}< 1$, 
for AP sending ${\left\langle g \right\rangle}_i, {\left\langle h \right\rangle}_i$ each time if $\frac{n_{I, i}\mu^2}{C(2\sigma_1^2+\sigma_2^2)} + \frac{n_{A, i}\mu^2}{C\sigma_2^2} \leq 2\left( \varepsilon_{\rm AP} - 2\ln\delta_{\rm AP} - 2\sqrt{\ln\delta_{\rm AP}(\ln\delta_{\rm AP}-\varepsilon_{\rm AP})} \right)$, where $\mu$ characterizes the value range of the gradients and Hessians.
\label{theorem:APLDP}
}

The proof can be found in Appendix \ref{APDPappendix}.
By \autoref{theorem:APLDP}, we observe that the noise energy required is related to the specific splitting vector $m_i$, enabling our algorithm to adaptively adjust the noise scale throughout the training process.
In the following section, we analyze the impact of $\sigma_1$ and $\sigma_2$ on PP's privacy leakage.

\subsection{Privacy Analysis: Differential Privacy at Passive Party Side} \label{PPprivacy}

\begin{figure*}[!tbp]
\centering
\includegraphics[width=180mm]{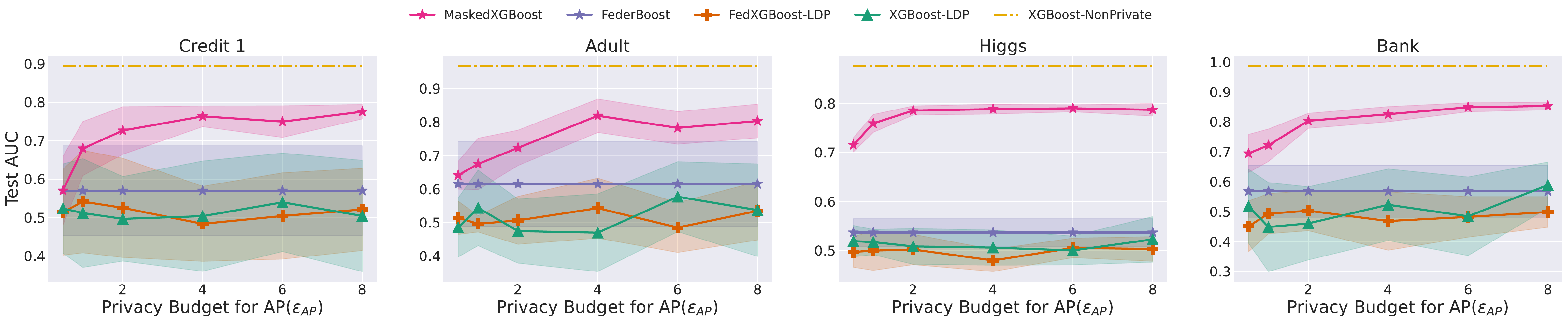}
\captionsetup{justification=centering}
\caption{Utility of different methods for different privacy budgets.}
\label{utilityfigure}
\end{figure*}

Upon receiving the noise matrices ${B_1,\ldots, B_l}$, privacy leakage occurs for PP.
Specifically, with a given noise vector $b_{ij}$, AP aims to infer $m_i$.
$m_i$ reveals the privacy of the user's feature data in PP. 
Recall Example \ref{example1}, where the feature vector is $d = [24, 25, 20, 22, 15, 17, 18, 16]^{\top}$, representing the ages of a group of people. 
A splitting candidate is chosen as Age = $18$. 
If AP can infer that $m_i = [0, 0, 0, 0, 1, 1, 1, 1]$, it can conclude that the ages of the first four individuals are greater than those of the last four, leading to a privacy disclosure of the users' age data.

Due to the designed disturbing noise in \autoref{alg:Secure-Noise}, AP is unable to deduce a potential solution for $m_i$ through straightforward brute force enumeration.
For each $m_i$, the output of \method is the noise matrix $B_i$. 
Since this represents aggregated information about $m_i$, we employ differential privacy (DP) to assess the privacy leakage of PP.

{\theorem{}
Recall that $n$ is the number of instances in datasets, $W$ is the number of noise vectors in a noise matrix $B_i$, i.e. $B_i = [b_{i1},\ldots,b_{iW}]$. 
For any $\varepsilon_{\rm PP}$, $\delta_{\rm PP}$ satisfy $\varepsilon_{\rm PP} > W\ln{2}$, $\delta_{\rm PP} \geq W {\bf Pr}_{y\sim\mathcal{N}(0,I_3)}\left(y^{\top}ky \geq 2\varepsilon_{\rm PP}/W - 2\ln2\right)$, \method is $(\varepsilon_{\rm PP}, \delta_{\rm PP})$-differentially private for PP sending each noise matrix $B_i$, where $k = diag(k_1,k_2,k_3)$, $k_1,k_2,k_3$ are the eigenvalues of 

$$
\begin{bmatrix}
    b-c & a+c & b-a\\
    a-b & 2b  & a-b\\
    b-a & a+c & b-c
\end{bmatrix}
$$

and 
$a = \frac{1}{n}\sum_{j=0}^{n-1} \frac{\sigma_1^2}{2\sigma_1^2[1-\cos(\frac{2\pi j}{n})]+\sigma_2^2}$, $b = \frac{1}{n}\sum_{j=0}^{n-1} \frac{\sigma_1^2\cos(\frac{2\pi j}{n})}{2\sigma_1^2[1-\cos(\frac{2\pi j}{n})]+\sigma_2^2}$ and $c = \frac{1}{n}\sum_{j=0}^{n-1} \frac{\sigma_1^2\cos(\frac{4\pi j}{n})}{2\sigma_1^2[1-\cos(\frac{2\pi j}{n})]+\sigma_2^2}$.
\label{theorem:PPDP}}

We provide detailed proof of \autoref{theorem:PPDP} in Appendix \ref{PPDPappendix}.
Based on \autoref{theorem:PPDP}, one can get a hint on the relation between noise ratio $\sigma_1^2/\sigma_2^2$ and privacy leakage in PP. 
According to our simulation results in Appendix \ref{PPDPappendix}, with fixed $\epsilon_{\rm PP}$, the lower-bound for $\delta_{\rm PP}$ increases monotonically with the noise ratio, indicating that a large value of $\sigma_1^2/\sigma_2^2$ translates to increased privacy leakage from PP. 
Intuitively, as the scale of disturbing noise $\sigma_2$ decreases, it becomes easier for AP to deduce the splitting vector by observing dimension additions close to zero, leveraging the insights from \autoref{alg:Secure-Noise}.

\begin{table}[!tbp] 
	\centering
        \vspace{0.2cm}
	\caption{Overview of the six datasets.}
    \setlength{\tabcolsep}{1.5mm}  
	\begin{tabular}{c|c c c c|c c c}  
        \toprule
	\textbf{Dataset} & \textbf{n.} & \textbf{f.} & \textbf{p.} &  \textbf{Dataset} & \textbf{n} & \textbf{f.} & \textbf{p.}\\ 
        \midrule
        Credit 1 & $150000$ & $10$ & $0.07$ & Higgs & $200000$ & $28$ & $0.47$ \\ 
        Adult & $32651$ & $14$ & $0.24$ & Bank & $45211$ & $16$ & $0.11$ \\ 
        Credit 2 & $30000$ & $23$ & $0.22$ & Nomao & $34465$ & $10$ & $0.28$ \\  
        \bottomrule
	\end{tabular}
 \label{datasettable}
\end{table}

\section{Empirical Evaluations} 
\label{experiments}
In this section, we first evaluate the model utility, convergence performance, and computational efficiency of \method, respectively. 
Secondly, we conduct experiments to show the empirical privacy-preserving property of \method. 
Thirdly, we conduct the ablation study to show how the PP privacy budget $\varepsilon_{\rm PP}$ affects \method's performance.

\subsection{Experimental Setup} \label{Experimental Setup}

\mypara{Datasets} 
We conduct our experiments on six widely used public XGBoost datasets including Credit 1, Adult, Higgs, Bank, Credit 2, and Nomao. 
All datasets are displayed in \autoref{datasettable} detailing the number of records($n.$), number of features($f.$), and proportion of the positive class ($p.$). 
All these real-world datasets are from Kaggle~\cite{kagglecredit,yeh2009comparisons} and the UCI repository~\cite{Dua:2019}.

\begin{figure*}[!tbp]
\centering
\includegraphics[width=180mm]{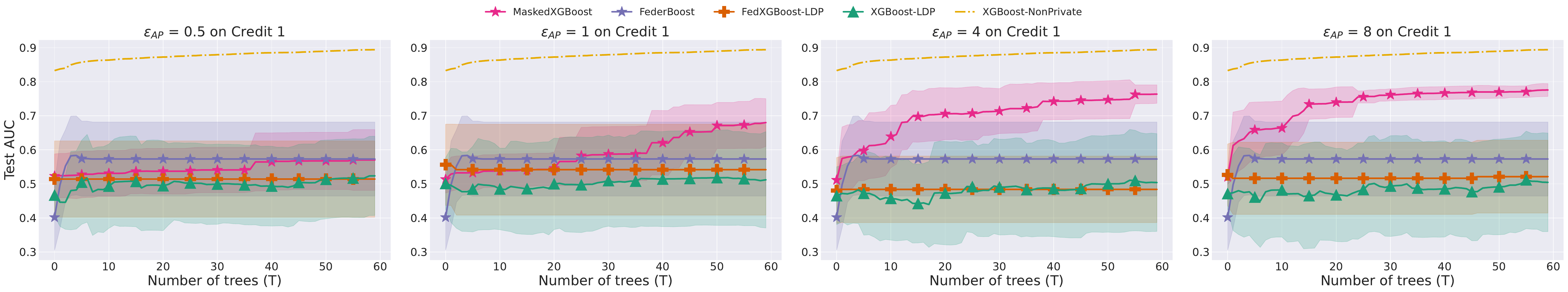} 
\caption{Training process of the four XGBoost algorithms with different privacy budgets on Credit 1 dataset} 
\label{trainprocesseps}
\end{figure*}

\mypara{Metrics} 
In the design of \method, we mainly consider two aspects of performance, model utility, and computational efficiency.

\begin{itemize}
    \item \mypara{(M1) Model Utility}
    We use the AUC\footnote{Measures such as accuracy are not useful for testing the performance of models in cases of class imbalance.} score to measure the model utility, which is widely used to evaluate the prediction ability of XGBoost models on binary classification. 
    \item \mypara{(M2) Computational Efficiency}
    We calculate the \textit{average training time per tree} to analyze the computational efficiency.
\end{itemize}

\mypara{Competitors} 
We have four baselines: (1) XGBoost-NonPrivate~\cite{chen2016}: Train XGBoost without privacy concerns. 
(2) XGBoost-LDP: Directly add Gaussian noise to gradients and Hessians to achieve LDP. 
(3) FedXGBoost-LDP~\cite{le2021fedxgboost}: Reduce the disturbance of LDP noises to the utility by modifying the splitting scoring function and utilizing only gradient information.
(4) FederBoost~\cite{tian2023sf}: Inject noise in data feature information of PP to achieve DP.
The definition of PP's privacy in this method is the same as in \method, that is, the relative order of private feature values of training data samples.

\mypara{Implementation}
We implement \method based on FedTree~\cite{fedtree}, which is an effective federated XGBoost template with C++. 
All experiments are run on a server containing 40 2.20 GHz CPUs, with 2 TB memory and Ubuntu 18.04 LTS system.
All experiments are repeated 10 times and the mean and standard deviation are reported.

\subsection{Utility Evaluation}
In this section, we evaluate the model utility of different methods. 
The metric is the test AUC with 60 training rounds and tree depth $d = 6$.
We choose $\varepsilon_{\rm AP}$ from $\{ 0.5,1,2,4,6,8 \}$. 
In all experiments we satisfy $(\varepsilon_{\rm AP},\delta_{\rm AP})$-LDP and $(\varepsilon_{\rm PP},\delta_{\rm PP})$-DP where $\delta_{\rm AP}=0.001$, and $\varepsilon_{\rm PP} = 1$, $\delta_{\rm PP} = 1/n$.
We use the advanced composition bound~\cite{dwork2010boosting} to calculate the privacy budget for each step.
XGBoost-NonPrivate represents the best AUC that can be achieved, and it is a horizontal line because its test AUC does not change with the privacy budget $\varepsilon_{\rm AP}$.
The evaluation results on four datasets are shown in \autoref{utilityfigure}, and the other results are in Appendix \ref{resultsappendix}.

\mypara{\method's Performance}
As illustrated in \autoref{utilityfigure}, we can clearly observe that \method is far superior to the other two methods. 
With the increment of privacy budget $\varepsilon_{\rm AP}$, the AUC of each method has been improved.
\method significantly outperforms the other baselines because a large portion of our noise used to protect privacy is carefully constructed and lossless.

\mypara{Other Approaches' Performance}
In \autoref{utilityfigure}, the AUCs of XGBoost-LDP and FedXGBoost-LDP have considerable fluctuations.  
During the XGBoost training process, if there is a significant error in the parent tree nodes, the error propagates to its left and right children nodes, resulting in larger errors.
Consequently, the AUCs for these methods exhibit notable instability and do not show substantial enhancement even with increased $\varepsilon_{\rm AP}$.

\method sometimes performs on average similarly to FederBoost when the AP privacy budget is low (e.g., $\varepsilon_{\rm AP} = 0.5$ on Credit 1 and Adult), but it is more stable. Moreover, \method significantly outperforms FederBoost as the AP privacy budget increases.

\subsection{Convergence Evaluation}

In order to better understand the differences between \method and other approaches, we analyze the training process of each algorithm. 
\autoref{trainprocesseps} depicts the training process of the four algorithms with different privacy budgets on Credit 1 dataset.
The results on other datasets are in Appendix \ref{resultsappendix}. 

\mypara{\method's Performance}
In general, we observe that the training process of the \method is stable, and the test AUC gradually rises.
As the privacy budget $\varepsilon_{\rm AP}$ grows, the AUC of \method is getting higher and closer to the ground truth: XGBoost-NonPrivate. 
As the privacy budget is relaxed, the randomness of different runs gradually decreases, demonstrating the stability of our algorithm.

\begin{table}[!tbp] 
	\centering 
        \vspace{0.2cm}
	\caption{Efficiency evaluation of different methods on six datasets.
    The average training time per tree is reported.
    We highlight our method in the \colorbox{pink}{red} ground.}
    \setlength{\tabcolsep}{0.5mm}
	\begin{tabular}{c|c c c c c c}  
        \toprule
	~ & \multicolumn{6}{c}{\textbf{Datasets}} \\
    \textbf{Methods} & Credit 1 & Adult & Higgs & Bank & Credit 2 & Nomao \\ 
        \midrule
        XGBoost-NonPrivate & $0.25$ & $0.15$ & $0.58$ & $0.17$ & $0.38$ & $0.15$ \\ 
        \method & $\cellcolor{pink}3.95$ & $\cellcolor{pink}1.54$ & $\cellcolor{pink}11.17$ & $\cellcolor{pink}2.21$ & $\cellcolor{pink}5.17$ & $\cellcolor{pink}1.77$\\
        HE-XGBoost & $25.22$ & $7.43$ & $47.70$ & $11.71$ & $18.41$ & $11.91$ \\
        XGBoost-LDP & $3.47$ & $1.41$ & $4.78$ & $1.29$ & $2.85$ & $1.45$\\
        FedXGBoost-LDP & $3.27$ & $1.18$ & $3.72$ & $1.23$ & $2.22$ & $1.03$\\
        FederBoost & $2.15$ & $1.11$ & $3.11$ & $1.05$ & $1.95$ & $0.95$ \\
        \bottomrule
	\end{tabular}
 \label{efficiency}
\end{table}

\begin{figure} [!tbp]
\centering  
\includegraphics[width=0.45\textwidth]{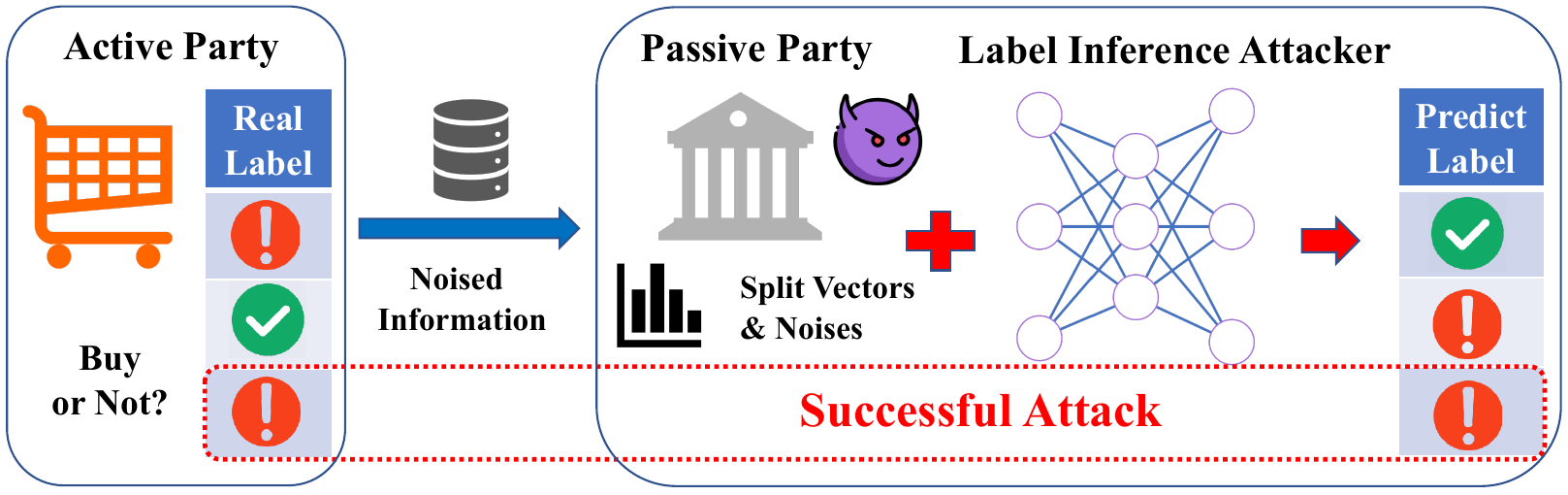}
\caption{Illustration of the neural network-based attacker.
PP receives the perturbed information and leverages the splitting vectors and noises in their possession to conduct a label inference attack using a pre-trained attacker model.}
\label{attackfigure}
\end{figure}


\mypara{Other Approaches' Performance}
The training process of the other three algorithms has a large disturbance, even if the number of training rounds increases, the performance cannot be improved stably. 
FedXGBoost-LDP has less disturbance than XGBoost-LDP due to the adjustment of the scoring function. 
However, since FedXGBoost-LDP does not consider the Hessian information, its performance is degraded.
The performance of XGBoost-LDP and FedXGBoost-LDP is not increased with the number of training rounds even with a relatively large privacy budget(i.e., $\varepsilon_{\rm AP}=8$).

Since there is no AP privacy breach in FederBoost, its training process are not affected by the AP privacy budget. 
However, the FederBoost AUC results are highly unstable due to significant perturbations in the feature data, which disrupt XGBoost's splitting candidate search process.

In XGBoost training, the error accumulates with the number of layers in a tree and also with the number of training rounds.
Therefore, increasing the number of training rounds (number of trees) does not guarantee an improvement in performance in the presence of noise.
\method adaptively adjusts the noise scale during the training process and increases communication to correct errors, resulting in reduced disturbance from noise during modeling.

\subsection{Efficiency Evaluation}
We show the training time comparison among \method, XGBoost-NonPrivate, XGBoost-LDP, FedXGBoost-LDP, FederBoost, and HE-XGBoost (using the HE method Paillier to protect privacy). 
The computation overhead of our approach mainly comes from building the categorical matrix $M$ and noise vectors. 
Thus, this overhead increases as the number of dimensions of training data increases. 
The majority of time consumption in HE-XGBoost arises from the encryption and decryption processes, as well as the cipher-text operations.
\autoref{efficiency} shows the average training time per tree of each method.
The training time per tree for \method is significantly lower than that of HE-XGBoost.
\method achieves $4.82 \times$ (Adult dataset) to $6.72 \times$ (Nomao dataset) training time improvement compared with the HE method.
Although our algorithm is not as efficient as other DP algorithms, it is acceptable, and our utility is better.
We provide the communication overhead analysis in Appendix \ref{appendix: communication}.

\begin{figure}[!tbp]
\centering
\includegraphics[width=0.48\textwidth]{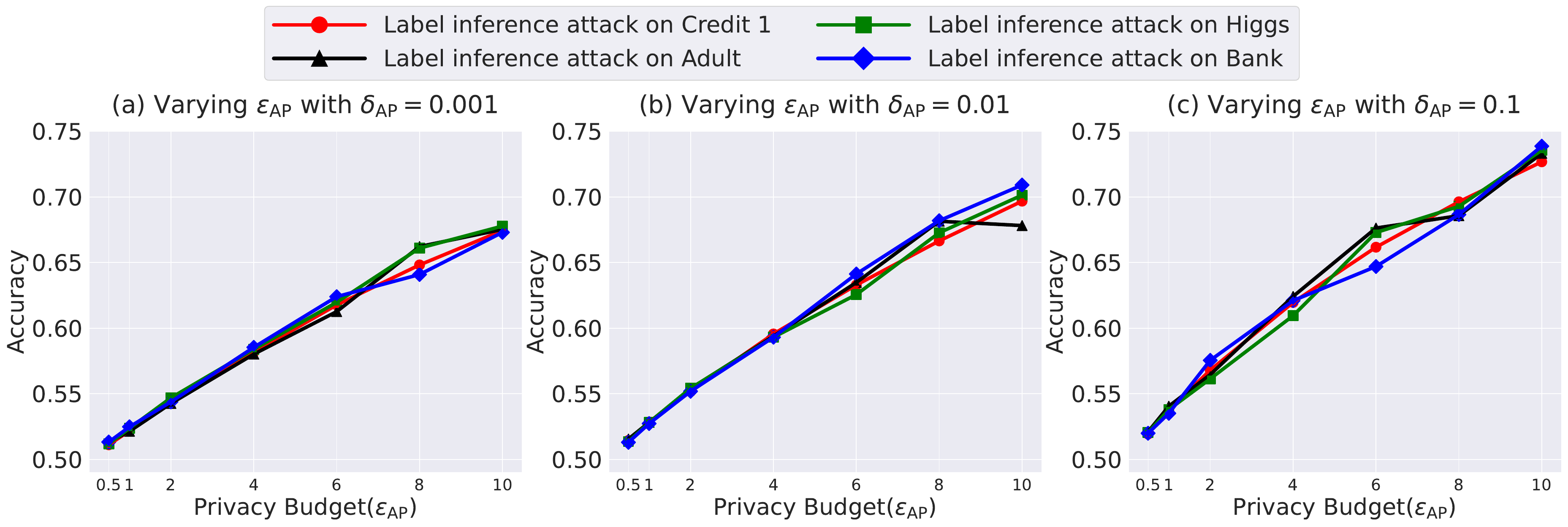}
\caption{Empirical privacy evaluation for AP with varying privacy budget $\varepsilon_{\rm AP}$ and $\delta_{\rm AP}$. 
As long as the privacy budget is small enough, the attacker cannot infer successfully.}
\label{1000attack}
\end{figure}

\subsection{Empirical Privacy Evaluation for AP}
\label{subsec:exp_attackforAP}

While we possess a theoretical privacy guarantee through local differential privacy, we also perform empirical privacy evaluations to confirm the adequacy of our algorithm's privacy.
In the honest-but-curious threat model, PP aims to perform a label inference attack. 
We introduce the following attacker for AP under this threat model.

\begin{figure}[!tbp]
\centering
\includegraphics[width=0.48\textwidth]{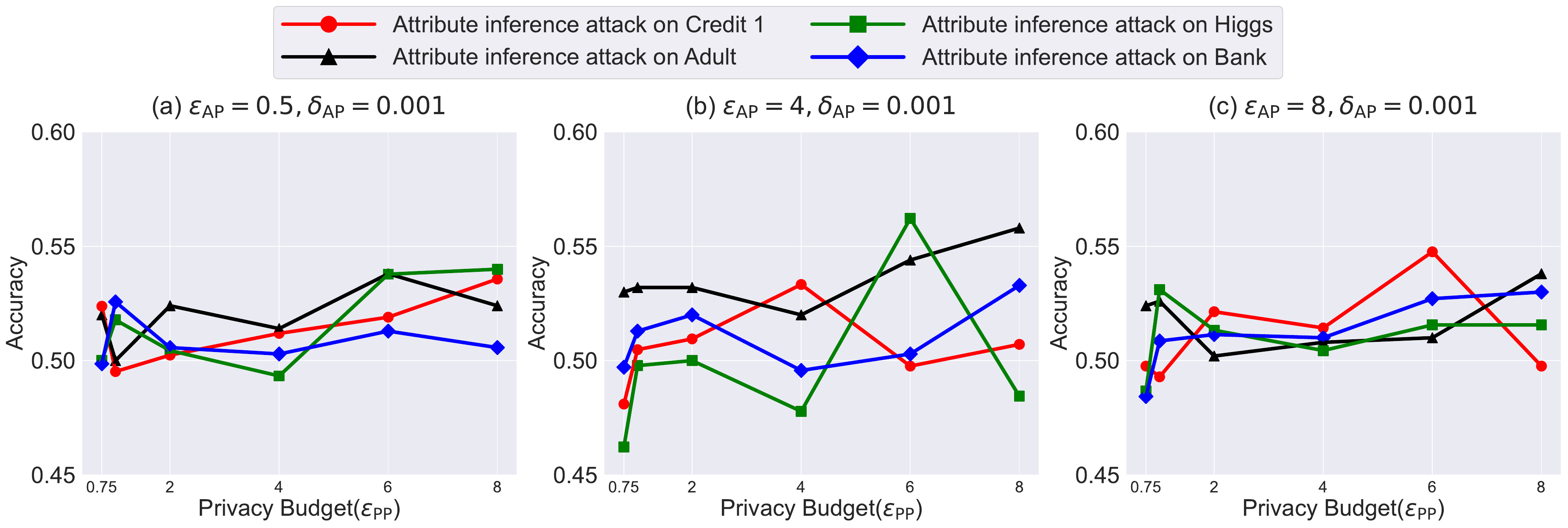}
\caption{Empirical privacy evaluation for PP with varying privacy budget $\varepsilon_{\rm PP}$ and $\varepsilon_{\rm AP}$.
Even with a very loose privacy budget(e.g., $\varepsilon_{\rm PP} = 8, \delta_{AP} = 8$), attacks on PP's splitting vector are still not effective.
}
\label{APattack_fourdatasets}
\end{figure}

\begin{figure*}[!tbp]
\centering
\includegraphics[width=180mm]{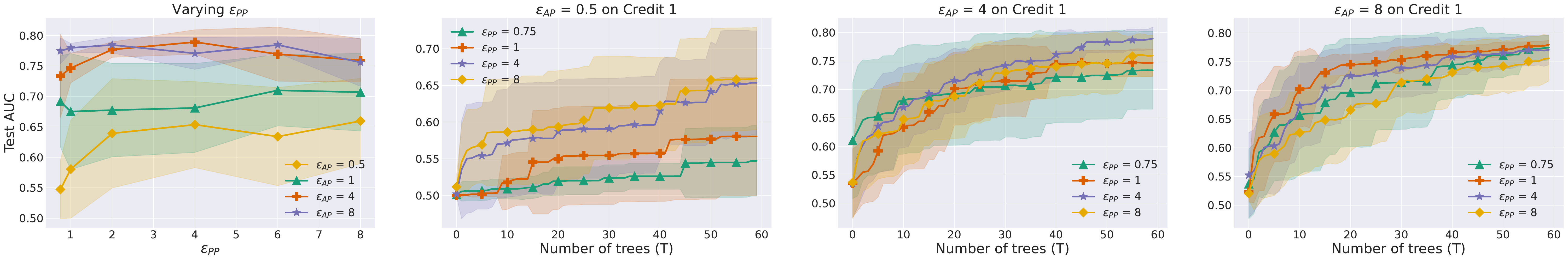}
\caption{Ablation study on Credit 1: the effect of $\varepsilon_{\rm PP}$ on final test AUC and training process.}
\label{ablationresult}
\end{figure*}

\mypara{Attack Setting \& Goal} 
PP builds the noise vectors $b_{i1},\ldots, b_{iW}$ according to a splitting vector $m_i$, and receive $\left\langle g \right\rangle _i,\left\langle h \right\rangle _i$ from AP, which are perturbed in virtue of ~\eqref{def::purtured_g_h}.
With the awareness of $b_{i1},\ldots, b_{iW}, {\left\langle g \right\rangle}_i, {\left\langle h \right\rangle}_i$, PP wants to de-noise ${\left\langle g \right\rangle}_i$ and ${\left\langle h \right\rangle}_i$. 
Then with the estimated plain information $g$ and $h$, PP can infer the label of instances. 
\autoref{attackfigure} shows the attack process of our attacker.

\mypara{Attack Methodology}
We construct a 5-layer neural network attacker for executing the attack.
This neural network provides estimates for the plaintext information, denoted as $g$ and $h$, which are subsequently used to calculate the predicted label.
PP emulates the real communication process by introducing noise and applying it to the fabricated gradients and Hessians. 
These fabricated datasets are generated based on PP's own splitting vector to simulate training data. 
For the testing, the attacker comprises the perturbed gradients and Hessians that AP sends during the actual execution of our method. 
We use L2 regularization with a parameter of 0.01 to mitigate overfitting.

\mypara{Evaluation Results}
After PP trains the attacker to convergence, we evaluate its performance by testing it on real, noisy gradients and Hessians, denoted as ${\left\langle g \right\rangle}_i$ and ${\left\langle h \right\rangle}_i$, which are generated by AP using our \method.
We conduct evaluations with different privacy budgets $\varepsilon_{\rm AP}$ and $\delta_{\rm AP}$ on six datasets.

In \autoref{1000attack}, we present the attack results on four datasets.
Other results are available in Appendix \ref{resultsappendix}.
Under a small privacy budget(e.g., $\varepsilon_{\rm AP} = 0.5, \delta_{AP} = 0.001$), the privacy of users can not be compromised.
When the privacy budget parameters are relaxed, the effectiveness of attacks slightly improves.

\subsection{Empirical Privacy Evaluation for PP}
\label{subsec:exp_attackforPP}

For the privacy of PP, besides providing theoretical differential privacy protection, we design an empirical privacy evaluation for PP.
In the honest-but-curious threat model, AP aims to perform an attribute inference attack. 
Under this threat model, we introduce the following attacker for PP.

\mypara{Attack Setting \& Goal}
AP receives the noise vectors $b_{i1},\ldots, b_{iW}$ according to the splitting vector $m_i$ owned by PP.
Since the distribution of these noise vectors is related to the splitting vector $m_i$, the attacker wants to infer the private splitting vector $m_i$ based on the noise vectors $b_{i1},\ldots, b_{iW}$.

\mypara{Attack Methodology}
Recall that noise consists of active noise, inactive noise, and disturbing noise in \autoref{subsec:noise_calibration}, where all dimensions of active noise add up to exactly $0$.
We assume that the attacker knows the size of the active set $n_{A,i}$.  
If no disturbing noise is injected, the attacker can traverse all possibilities $C^{n_{A,i}}_n$, where the one that adds exactly $0$ is the correct active set.
Due to the addition of disturbance noise, the corresponding dimensions of the active set are not added to $0$, but the corresponding disturbance is still much smaller than that of other combinations. 
So the attacker can find the predicted active set based on the minimum absolute value of the sum.

\mypara{Evaluation Results}
We measure the attack effectiveness by the bit match between the splitting vector predicted by the attacker and the real vector.
We conduct evaluations with different privacy budgets $\varepsilon_{\rm PP}$ and $\varepsilon_{\rm AP}$ and on six datasets.
$\delta_{\rm PP}$ and $\delta_{\rm AP}$ is set to be $1/n,0.001$ respectively.

In \autoref{APattack_fourdatasets}, we present the attack results on four datasets.
Other results are available in Appendix \ref{resultsappendix}.
We find that the attack is ineffective due to the presence of disturbing noise, and even when the privacy budget is relaxed, the attack success rate does not exceed $0.6$. This demonstrates that \method fully guarantees the privacy of PP.
The empirical privacy evaluation is based on the honest-but-curious adversary assumption, we provide the analysis if adversaries deviate from the protocol in Appendix \ref{appendix: adversary}.

\subsection{Ablation Study} 
\label{ablation}

Now we evaluate the impact of $\varepsilon_{\rm PP}$ on \method's performance. 
As shown in \autoref{ablationresult}, \method's utility improves with higher $\varepsilon_{\rm PP}$ values. 
This aligns with the findings in \autoref{theorem:utility} and \autoref{theorem:PPDP}, where a larger $\varepsilon_{\rm PP}$ corresponds to a smaller $\sigma_2$, allowing \method to improve performance.
Consider the training process, regardless of the $\varepsilon_{\rm PP}$ setting, \method exhibits a relatively fast convergence rate, although there may be a slight difference in the final test AUC. 
For other datasets' results, please refer to Appendix \ref{resultsappendix}.

\section{Related Work}
\label{sec:related}

\mypara{Vertical Federated XGBoost}
Recently, Cheng et al.~\cite{cheng2021secureboost} proposed SecureBoost, a federated XGBoost framework for vertically partitioned data. 
In SecureBoost, AP calculates $g_i$ and $h_i$ for all samples, encrypts them using additively homomorphic encryption, and sends the ciphertexts to PP.
Using the homomorphic property, PP computes the ciphertext sum of gradients and Hessians of left and right nodes for all possible splitting candidates and returns the ciphertexts to AP. 
AP decrypts all ciphertexts and finds the best split. 
This protocol is expensive since it requires cryptographic computation and communication for each possible split. 

Some related works propose approaches that avoid encryption operations to get better efficiency. 
Tian et al.~\cite{tian2023sf} suppose that PP adds differentially private noise to the users' bucket, which is formed by the division of splitting candidates.
Such an approach is still impractical because the user information in a bucket is highly compromised and performance decreases significantly as the privacy budget decreases.

\mypara{Horizontal Federated XGBoost}
Recent studies have explored GBDT implementations, such as XGBoost, for secure training in horizontal federated settings~\cite{deforth2022xorboost, feng2019securegbm}. 
These approaches typically rely on cryptographic techniques like Secure Multi-Party Computation (MPC) or HE, which result in high computational and communication overhead. 
Tian et al.~\cite{tian2023sf} propose methods under the local DP model but experience significant utility loss due to the local noise\cite{tian2023sf}. 
Maddock et al.~\cite{maddock2022hltreewithdp} use lightweight MPC techniques such as secure aggregation to balance DP efficiency with cryptographic security. 
Peinemann et al.~\cite{peinemann2023s} introduce non-spherical multivariate Gaussian noise to improve the utility-privacy trade-off.

\section{Conclusion and Future Work} 
\label{conclusion}
In this paper, we proposed \method that enables the state-of-the-art tree ensemble model XGBoost to be conducted under VFL settings. 
Different from the previous work applying homomorphic encryption, our protocol \method achieves lower overhead while maintaining consistently better accuracy and provides a bilateral differential privacy guarantee with rigorous proof. 
In our protocol \method, we design a special privacy protection mechanism for VFL XGBoost instead of directly applying DP noise perturbation, which may have great potential to be applied to other privacy-preserving algorithms.

In our future work, we plan to enhance the performance of \method. This involves optimizing the construction of the categorical matrix and the matrix operation process. Additionally, we aim to implement \method in real-world industrial scenarios to evaluate its performance with more realistic data.

\bibliographystyle{ACM-Reference-Format}
\bibliography{maskedxgb}

\appendix

\section{Details of XGBoost}\label{xgboostappendix}

\mypara{Regression Tree}
A regression tree, which belongs to the family of decision trees, is a commonly used algorithm for predicting numerical output variables in a given dataset, where the numerical output variables are defined as
\[  f(x) = w_{q(x)} , q: \mathcal{X} \xrightarrow{} \left\{ 1,\ldots,L \right\}, w \in \mathbb{R}^{L}  \]
where $q$ denotes the tree structure that maps the feature of an instance to a unique leaf, $w$ is the weight vector of the leaf, and $L$ is the number of leaves of one tree.

The regression tree is capable of assigning any instance $x$ to a specific leaf node with a leaf index $q(x)$. This index corresponds to a unique leaf weight value $w_{q(x)}$. 
The process of training a regression tree is essentially learning a \textit{tree structure} $q$ and \textit{leaf weight} $w$.

\mypara{Splitting Rules}
To learn a tree structure at the $t^{th}$ iteration, many \textit{splitting rules}, i.e., mappings that assign an instance $x_i$ to a specific leaf node, are proposed according to the order of feature data~\cite{friedman2001greedy}.  
The algorithm proposed in~\cite{chen2016} is a greedy rule that splits the instances into two disjoint sets of instances (which are associated with the left and right child nodes) at a node and repeats the splitting multiple times. 
\eqref{eqn: LoptimalSplit} is often used to evaluate a splitting candidate at each node $j$ (associated with instance set $\mathcal I_j$) ~\cite{chen2016}.
In \eqref{eqn: LoptimalSplit}, $\mathcal I_{L,j}$ and $ \mathcal I_{R,j}$ are associated with the left and right child nodes, forming the dichotomy of the proposed splitting in question, and  $\lambda$ and $\gamma$ are the regularization parameters in \eqref{eqn:costFuncL}.

The proposed that gets the highest score in \eqref{eqn: LoptimalSplit} will be selected as the tree structure at this node.
Then, splitting continues from the newly constructed nodes until it reaches the maximum depth of the tree. 
The deepest nodes are called the leaves of a tree.

\section{Proof of Theorem~\ref{theorem:utility}}
\label{losslessappendix}

We use $L_{split}^{s_i}$ to denote the true splitting score of the $i^{th}$ splitting candidate and $\langle L_{split}^{s_i}\rangle $ to denote \method's estimation of it. 
$m_i$ is the corresponding splitting vector.
${\left\langle g \right\rangle}_i, {\left\langle h \right\rangle}_i  \in \mathbb R^n$ are the noised gradient and Hessian vectors of the $n$ instances.
$G$ and $H$ are the sums of all gradients and hessians.
We have

$$
\begin{aligned}
L_{split}^{s_i} = - \gamma + \frac{1}{2}\left( \frac{(g^L_{i})^2}{(h^L_{i}) + \lambda} + \frac{(g^R_{i})^2}{(h^R_{i}) + \lambda} - \frac{G^2}{H + \lambda} \right). \\
\langle L_{split}^{s_i}\rangle = - \gamma + \frac{1}{2}\left( \frac{({\langle g^L_i \rangle})^2}{{\langle h^L_i \rangle} + \lambda} + \frac{({\langle g^R_i \rangle})^2}{{\langle h^R_i \rangle} + \lambda}  - \frac{G^2}{H + \lambda} \right).
\end{aligned}
$$
where $\langle g^L_i \rangle = m_i^{\top}{\left\langle g \right\rangle}_i, \langle h^L_i \rangle = m_i^{\top}{\left\langle h \right\rangle}_i, \langle g^R_i \rangle = G - \langle g^L_i \rangle, \langle h^R_i \rangle = H - \langle h^L_i \rangle$. 
According to \autoref{subsec:noise_calibration}, we have $ \mathbb{E}[\langle g^L_i \rangle] = g^L_{i}, Var[\langle g^L_i \rangle] = Var[\langle g^R_i \rangle]  = n_AC\sigma_2^2$, and we denote the variance by $\kappa^2$ in \autoref{theorem:utility}.

We want to upper bound:
\begin{equation}\label{eqn:prob_bound1}
	{\bf Pr}\left(\mid \langle L_{split}^{s_i}\rangle  - L_{split}^{s_i} \mid \geq \alpha \right)\quad \forall\alpha>0
\end{equation}

Denote $X:=\langle g_i^L \rangle$, $Y:=\langle h_i^L \rangle+\lambda$, $Z:=\langle g_i^R \rangle$
and $W:=\langle h_i^R \rangle+\lambda$, then $X,Y,Z,W$ are normal with variance $\kappa^2$ and mean $\mu_X,\mu_Y,\mu_Z,\mu_W$. Denote $C_L:=\frac{(g_i^L)^2}{(h_i^L)+\lambda}$ and $C_R:=\frac{(g_i^R)^2}{(h_i^R)+\lambda}$. We can write~\eqref{eqn:prob_bound1} as:
\begin{equation}\label{eqn:prob_bound2}
	{\bf Pr}\left(\mid \frac{X^2}{Y}-C_L+\frac{Z^2}{W}-C_R\mid \geq 2\alpha\right).
\end{equation}
We have~\eqref{eqn:prob_bound2} upper bounded by
${\bf Pr}(L)+{\bf Pr}(R)$, as short hands for ${\bf Pr}\left(\mid \frac{X^2}{Y}- C_L\mid\geq\alpha\right)$ and ${\bf Pr}\left(\mid\frac{Z^2}{W}-C_R\mid \geq \alpha\right)$, respectively. It's also possible to derive a tighter bound ${\bf Pr}(L)+{\bf Pr}(R)-{\bf Pr}(L\cap R)$, but much more tedious.

For simplicity, we only present the derivation of ${\bf Pr}(L)$ while omit the derivation for ${\bf Pr}(R)$, which is very similar. Denote $C_L+\alpha:=\beta_L$, $C_L-\alpha:=\beta_L'$, we proceed by conditioning $L$ on the sign of $Y$:
\vspace{-3pt}
\begin{align}
{\bf Pr}(L)&={\bf Pr}\left(A\cap\{Y>0\}\right)+{\bf Pr}\left(B\cap\{Y>0\}\right)\nonumber\\
&+{\bf Pr}\left(B^c\cap\{Y<0\}\right)+{\bf Pr}\left(A^c\cap\{Y<0\}\right)\label{eqn:PL_decomposition}
\end{align}
where $A:=\{X^2\geq
\beta_L Y\}$, $B:=\{X^2\leq\beta_L'Y\}$.\\

We can write the first term in~\eqref{eqn:PL_decomposition} as an integration: 
\begin{align*}
	&\int_{0}^{\infty} {\bf Pr}\left(X^2\geq\beta_Lt\right){\rm d}F_Y(t) \\
	=&\int_{0}^{\infty} \left(\Phi\left(\frac{-\sqrt{\beta_Lt}+\mu_X}{\kappa}\right)+\Phi\left(\frac{-\sqrt{\beta_Lt}-\mu_X}{\kappa}\right)\right){\rm d}F_Y(t)
\end{align*}
where ${\rm d}F_Y(t)=\frac{1}{\kappa}\phi\left(\frac{t-\mu_Y}{\kappa}\right){\rm d}t$, $\phi(\cdot)$ and $\Phi(\cdot)$ are the density and cdf of the standard normal distribution.

Similar integration can be done for the second and third terms in~\eqref{eqn:PL_decomposition}, while the last term is zero. In summary:
\begin{itemize}
\item If $\alpha\leq C_L$, i.e. $\beta_L'\geq 0$, ${\bf Pr}(L)$ is no greater than:
\begin{equation}\label{eqn:left_upperbound_1}
	{\bf Pr}\left(A\cap\{Y>0\}\right)+{\bf Pr}\left(B\cap\{Y>0\}\right)+{\bf Pr}\left(\{Y<0\}\right)
\end{equation}
\item if $\alpha\geq C_L$, i.e. $\beta_L'\leq 0$, ${\bf Pr}(L)$ is no greater than:
\begin{equation}\label{eqn:left_upperbound_2}
	{\bf Pr}\left(A\cap\{Y>0\}\right)+{\bf Pr}\left(B^c\cap\{Y<0\}\right)
\end{equation}
\end{itemize}

As $\kappa^2\downarrow0$, $Y\stackrel{a.s.}{\rightarrow}\mu_Y$, $X^2\stackrel{a.s.}{\rightarrow}\mu_X^2$. From the facts that $C_L\cdot\mu_Y=\mu_X^2$ and $\mu_Y>0$, we conclude that as $\kappa^2\downarrow0$, ${\bf Pr}(A)$, ${\bf Pr}(B)$, and ${\bf Pr}(\{Y<0\})$ converge to zero; as a result, the upper bounds~\eqref{eqn:left_upperbound_1} and~\eqref{eqn:left_upperbound_2} converge to zero.\hfill $\blacksquare$
{\remark To facilitate numerical evaluation, we can further use various inequalities~\cite{duembgen2010bounding} for $\Phi(\cdot)$ in integration, such as:
$$
	\frac{2\phi(x)}{\sqrt{4+x^2}+x}<1-\Phi(x)<\frac{2\phi(x)}{\sqrt{8/\pi+x^2}+x}\quad\forall x>0
$$
}

\section{Proof of Theorem~\ref{theorem:APLDP}}
\label{APDPappendix}


Assume $g, g' \in \mathbb{R}^n$ are any two gradient vectors of $n$ instances from AP, $g = [g_1, \ldots, g_n]^{\top}, g' = [g'_1, \ldots, g'_n]^{\top}$.
Let $v = g - g' = [v_1, \ldots, v_n]$, and we have $|v_i| \leq \mu$.
For each splitting vector $m_i$, our algorithm is equivalent to adding a noise vector on $g$: ${\left\langle g \right\rangle}_i = g + \eta_i, \eta_i \sim \mathcal{N}(0, \Sigma)$, where the submatrix with rows and columns indexed by $\mathcal{A}(m_i)$ is given as
\begin{align}
&\Sigma_{2} = \\
&\begin{bmatrix}
C(2\sigma_1^2 + \sigma_2^2) & -C\sigma_1^2 & 0& \cdots& 0 & -C\sigma_1^2 \\
-C\sigma_1^2 &\ddots & \ddots & \ddots & & 0\\
0 & & \ddots & \ddots & \ddots&  \vdots\\
\vdots& & &\ddots & \ddots & 0\\
0 & & & & \ddots& -C\sigma_1^2 \\
-C\sigma_1^2 & 0 &\cdots & 0& -C\sigma_1^2 & C(2\sigma_1^2 + \sigma_2^2)
\end{bmatrix}\notag
\end{align}
and the submatrix of all the other entries is given as
\begin{equation}
\Sigma_{1} = C(2\sigma_1^2+\sigma_2^2)I.
\end{equation}

Notice that $\Sigma_{1} \in \mathbb{R}^{n_{I,i} \times n_{I,i}}, \Sigma_{2} \in \mathbb{R}^{n_{A,i} \times n_{A,i}}$, where $n_{I,i} = |\mathcal{I}(m_i)|, n_{A,i} = |\mathcal{A}(m_i)|$.
Without loss of generality, we assume $\Sigma = diag(\Sigma_{1}, \Sigma_{2})$.
For any measurable set $\mathcal{O} \subset \mathbb{R}^n$, we have
$$
\begin{aligned}
& {\bf Pr} ({\left\langle g \right\rangle}_i \in \mathcal{O}) \\
= & \int_{\mathbb{R}^n} \frac{1}{\sqrt{(2\pi)^n|\Sigma|}} 1_\mathcal{O}(g + w) e^{-\frac{1}{2}w^{\top}\Sigma^{-1}w } {\rm d} w \\
= & \int_{\mathbb{R}^n} \frac{1}{\sqrt{(2\pi)^n|\Sigma|}} 1_\mathcal{O}(u) e^{-\frac{1}{2}(u-g)^{\top}\Sigma^{-1}(u-g) } {\rm d} u \\
= & \int_{\mathcal{O}} \frac{1}{\sqrt{(2\pi)^n|\Sigma|}} e^{-\frac{1}{2}(u-g')^{\top}\Sigma^{-1}(u-g') } e^{(u-g')^{\top}\Sigma^{-1}v - \frac{1}{2}v^{\top}\Sigma^{-1}v} {\rm d} u \\
\leq & \int_{\mathcal{O}} \frac{1}{\sqrt{(2\pi)^n|\Sigma|}} e^{-\frac{1}{2}(u-g)^{\top}\Sigma^{-1}(u-g) } 1_\mathcal{\mathcal{S}}(u)  {\rm d} u  \\ 
& + \exp(\varepsilon_{\rm AP}){\bf Pr} ({\left\langle g' \right\rangle}_i \in \mathcal{O}),
\end{aligned}
$$

where $\mathcal{S} := \{ u \in \mathbb{R}^n| e^{(u-g')^{\top}\Sigma^{-1}v - \frac{1}{2}v^{\top}\Sigma^{-1}v} \geq e^{\varepsilon_{\rm AP}}\}$.
Now we proceed to bound the integral term.
$$
\begin{aligned}
& \int_{\mathcal{O}} \frac{1}{\sqrt{(2\pi)^n|\Sigma|}} e^{-\frac{1}{2}(u-g)^{\top}\Sigma^{-1}(u-g) } 1_\mathcal{\mathcal{S}}(u)  {\rm d} u \\
\leq & \int_{\mathbb{R}^n} \frac{1}{\sqrt{(2\pi)^n|\Sigma|}} e^{-\frac{1}{2}(u-g)^{\top}\Sigma^{-1}(u-g) } 1_\mathcal{\mathcal{S}}(u)  {\rm d} u \\
= & {\bf Pr} ((u-g')^{\top}\Sigma^{-1}v - \frac{1}{2}v^{\top}\Sigma^{-1}v \geq \varepsilon_{\rm AP}) \\
= & {\bf Pr}_{w \sim \mathcal{N}(0, \Sigma)} ((w+v)^{\top}\Sigma^{-1}v - \frac{1}{2}v^{\top}\Sigma^{-1}v \geq \varepsilon_{\rm AP} ) \\
= & {\bf Pr}_{w \sim \mathcal{N}(0, \Sigma)} (w^{\top}\Sigma^{-1}v + \frac{1}{2}v^{\top}\Sigma^{-1}v \geq \varepsilon_{\rm AP}) \\
\end{aligned}
$$

We denote $v = [v_1^{\top}, v_2^{\top}], v_1 \in \mathbb{R}^{n_{I,i}}, v_2 \in \mathbb{R}^{n_{A,i}}$, then we bound the above term as follows:

$$
\begin{aligned} 
& {\bf Pr} (y^{\top}\Sigma^{-\frac{1}{2}}v + \frac{1}{2}(v_1^{\top}\Sigma_1^{-1}v_1 + v_2^{\top}\Sigma_2^{-1}v_2)  \geq \varepsilon_{\rm AP}) \\
\leq & {\bf Pr} (y^{\top}\Sigma^{-\frac{1}{2}}v \geq \varepsilon_{\rm AP} - \frac{{n_{I,i}} \mu^2}{2C(2\sigma_1^2 + \sigma_2^2)} - \frac{{n_{A,i}} \mu^2}{2C\sigma_2^2})
\end{aligned}
$$
where $y \sim \mathcal{N}(0, I_{n})$, $I_{n}$ is the $n$-by-$n$ identity matrix.
The inequality follows from the eigenvalues bound of $\Sigma_2^{-1}$~\cite{tee2007eigenvectors_circulant}:
$$
\lambda_j=\frac{1}{C(2\sigma_1^2[1-\cos(\frac{2\pi j}{n})]+\sigma_2^2)} \leq \frac{1}{C\sigma_2^2}
$$
Notice that $y^{\top}\Sigma^{-\frac{1}{2}}v \sim \mathcal{N}(0, \norm{\Sigma^{-\frac{1}{2}}v}_2^2)$, and 
$$
\begin{gathered}
\norm{\Sigma^{-\frac{1}{2}}v}_2^2  = v^{\top}\Sigma^{-1}v \leq \frac{{n_{I,i}} \mu^2}{C(2\sigma_1^2 + \sigma_2^2)} + \frac{{n_{A,i}} \mu^2}{C\sigma_2^2}
\end{gathered}
$$

Let $\sigma^2 = \frac{{n_{I,i}} \mu^2}{C(2\sigma_1^2 + \sigma_2^2)} + \frac{{n_{A,i}} \mu^2}{C\sigma_2^2}$, we have 
$$
\begin{aligned}
{\bf Pr} (y^{\top}\Sigma^{-\frac{1}{2}}v \geq \varepsilon_{\rm AP} - \frac{\sigma^2}{2}) \leq  {\bf Pr}_{z \sim \mathcal{N}(0, \sigma^2)} (z \geq \varepsilon_{\rm AP} - \frac{\sigma^2}{2})
\end{aligned}
$$

If $\sigma^2 \leq 2\left( \varepsilon_{\rm AP} - 2\ln\delta_{\rm AP} - 2\sqrt{\ln\delta_{\rm AP}(\ln\delta_{\rm AP}-\varepsilon_{\rm AP})} \right)$, we can verify that

$$
\begin{gathered}
{\bf Pr}_{z \sim \mathcal{N}(0,\sigma^2)} (z \geq \frac{\sigma^2}{2}) \leq \delta_{\rm AP}
\end{gathered}
$$

where the last inequality follows from a Gaussian tail bound.

\section{Proof of Theorem~\ref{theorem:PPDP}}
\label{PPDPappendix}

Without loss of generality, we consider two adjacent splitting vectors $m_i,m_i^{\prime} $ that only differ in the first entry (i.e.,$m_{i1} = 0, m^{\prime}_{i1} = 1$) and with all other entries being 1.
The other cases can be proved similarly. With $m_i$ and $m_i^{\prime}$, 
PP generates two noise matrices, denoted as $B_i$ and $B_i'$. Each 
column vector $b_{ij}$($j=1,\ldots,W$) of the matrix $B_i$ is randomly and independently drawn from
$\mathcal{N}\left(0, \Sigma_1\right)$ and 
each column vector $b_{ij}'$($j=1,\ldots,W$) of the matrix $B'$ randomly and independently drawn from
$ \mathcal{N}\left(0, \Sigma_2\right)$, where 
$\Sigma_1$ and 
$\Sigma_2$ are shown as follows:
$$  
\Sigma_2 = 
\begin{bmatrix}
2\sigma_1^2 + \sigma_2^2 & -\sigma_1^2 & 0& \cdots& 0 & -\sigma_1^2 \\
-\sigma_1^2 &\ddots & \ddots & \ddots & & 0\\
0 & & \ddots & \ddots & \ddots&  \vdots\\
\vdots& & &\ddots & \ddots & 0\\
0 & & & & \ddots& -\sigma_1^2 \\
-\sigma_1^2 & 0 &\cdots & 0& -\sigma_1^2 & 2\sigma_1^2 + \sigma_2^2
\end{bmatrix}
$$

$$
\begin{aligned}
& \Sigma_1 = \\
& 
\begin{bmatrix}
\begin{array}{c:cccccc}
2\sigma_1^2 + \sigma_2^2& 0 & 0 & 0 &\cdots & 0 & 0 \\ \hdashline
0&2\sigma_1^2 + \sigma_2^2 & -\sigma_1^2 & 0& \cdots& 0 & -\sigma_1^2 \\
0&-\sigma_1^2 &\ddots & \ddots & \ddots & & 0\\
0&0 & & \ddots & \ddots & \ddots&  \vdots\\
\vdots& \vdots& & & \ddots & \ddots & 0\\
0&0 & & & & \ddots& -\sigma_1^2 \\
0&-\sigma_1^2 & 0 &\cdots & 0& -\sigma_1^2 & 2\sigma_1^2 + \sigma_2^2
\end{array}
\end{bmatrix}
\end{aligned}
$$

Since each column of $B_i$ is generated independently, we can only consider $b_{i1}$ and $b_{i1}'$ and apply Lemma~\ref{DPsequential} to obtain the result for the whole matrix. 

Denote $\Sigma :=\Sigma_1^{-1} - \Sigma_2^{-1}$ and $\mathcal S:=\{w:\sqrt{\frac{|\Sigma_2|}{|\Sigma_1|}} e^{-\frac{1}{2}w^{\top} \Sigma w} \geq e^\varepsilon\}$, for any measurable set $\mathcal{O} \subset \mathbb{R}^n$, we have
$$
\begin{aligned}
& {\bf Pr} (b_{i1} \in \mathcal{O}) = \int_{\mathcal{O}} \frac{1}{\sqrt{(2\pi)^n|\Sigma_1|}} e^{-\frac{1}{2}w^{\top}\Sigma_1^{-1}w } {\rm d} w \\
= & \int_{\mathcal{O}} \frac{1}{\sqrt{(2\pi)^n|\Sigma_1|}} e^{-\frac{1}{2}w^{\top}\Sigma_2^{-1}w} e^{-\frac{1}{2}w^{\top} \Sigma w}{\rm d} w \\
= & \int_\mathcal{O} \frac{1}{\sqrt{(2\pi)^n|\Sigma_2|}} e^{-\frac{1}{2}w^{\top}\Sigma_2^{-1}w} \sqrt{\frac{|\Sigma_2|}{|\Sigma_1|}} e^{-\frac{1}{2}w^{\top} \Sigma w}{\rm d} w  \\
\leq & \int_\mathcal{O} \frac{1}{\sqrt{(2\pi)^n|\Sigma_1|}} e^{-\frac{1}{2}w^{\top}\Sigma_1^{-1}w} 1_{\mathcal S}(w){\rm d} w \\
& + \exp(\varepsilon){\bf Pr} (b_{i1}' \in \mathcal{O}),
\end{aligned}
$$
Now we proceed to bound the integral term. 
The following fact is used in the derivation: the determinant ratio $\frac{|\Sigma_2|}{|\Sigma_1|} \in (\frac{1}{4},2)$, and we provide the proof of it in Appendix \ref{proof:detbound}.
$$
\begin{aligned}
 & \int_\mathcal{O} \frac{1}{\sqrt{(2\pi)^n|\Sigma_1|}} e^{-\frac{1}{2}w^{\top}\Sigma_1^{-1}w} 1_{\mathcal S}(w) {\rm d} w \\
 \leq & \int_{\mathbb{R}^n} \frac{1}{\sqrt{(2\pi)^n|\Sigma_1|}} e^{-\frac{1}{2}w^{\top}\Sigma_1^{-1}w} 1_{\mathcal S}(w) {\rm d} w  \\
 = & {\bf Pr}_{w\sim\mathcal{N}(0,\Sigma_1)}\left( \sqrt{\frac{|\Sigma_2|}{|\Sigma_1|}} e^{-\frac{1}{2}w^{\top} \Sigma w} \geq e^\varepsilon \right) \\
 < & {\bf Pr}_{w\sim\mathcal{N}(0,\Sigma_1)}\left({2}e^{-\frac{1}{2}w^{\top} \Sigma w} \geq e^\varepsilon \right) \\
 = & {\bf Pr}_{w\sim\mathcal{N}(0,\Sigma_1)} \left( w^{\top}(-\Sigma)w \geq 2\varepsilon- 2\ln 2 \right) \\ 
 = & {\bf Pr}_{y\sim\mathcal{N}(0,I)}\left( y^{\top}\Sigma_1^{\frac{1}{2}}(\Sigma_2^{-1} - \Sigma_1^{-1})\Sigma_1^{\frac{1}{2}}y \geq 2\varepsilon- 2\ln 2 \right) \\
 = & {\bf Pr}_{y\sim\mathcal{N}(0,I)} \left( y^{\top}(\Sigma_1^{\frac{1}{2}}\Sigma_2^{-1}\Sigma_1^{\frac{1}{2}} - I)y \geq 2\varepsilon- 2\ln 2 \right)
\end{aligned} 
$$
Because $\Sigma_1^{\frac{1}{2}}\Sigma_2^{-1}\Sigma_1^{\frac{1}{2}} - I$ is symmetric, we have unitary decomposition 
$$
\Sigma_1^{\frac{1}{2}}\Sigma_2^{-1}\Sigma_1^{\frac{1}{2}} - I=U\Lambda U^{\top}
$$
where $U$ is orthogonal, and $\Lambda$ is a diagonal matrix composed of the eigenvalues of $\Sigma_1^{\frac{1}{2}}\Sigma_2^{-1}\Sigma_1^{\frac{1}{2}} - I$, where $y\sim\mathcal{N}(0,I)$, we have $U^{\top}y\sim\mathcal{N}(0,I)$. It follows that 
$$
\begin{aligned}    
& {\bf Pr}_{y\sim\mathcal{N}(0,I)}\left( y^{\top}(\Sigma_1^{\frac{1}{2}}\Sigma_2^{-1}\Sigma_1^{\frac{1}{2}} - I)y \geq 2\varepsilon- \ln 2 \right) \\
= & {\bf Pr}_{y\sim\mathcal{N}(0,I)}\left( y^{\top}\Lambda y \geq 2\varepsilon- 2\ln 2 \right)
\end{aligned}
$$

Next, we prove that $\Lambda$ only has three non-zero diagonals. 
By Theorem 1.3.22 of~\cite{horn2012matrix}, the set of the eigenvalues of $\Sigma_1^{\frac{1}{2}}\Sigma_2^{-1}\Sigma_1^{\frac{1}{2}} - I$ are equal to that of $\Sigma_2^{-1}\Sigma_1- I$, which are further equal to that of $(P\Sigma_2P^{\top})^{-1}P(\Sigma_1-\Sigma_2)P^{\top}$, where $P$ can be any orthogonal matrix.
Specifically, we choose the following permutation matrix:
$$
P =\begin{bmatrix}
0&0 & \cdots & 0 &1 \\
1 & 0 & \cdots & \cdots & 0\\
0 & 1 & 0 & \cdots & 0\\
\vdots & & \ddots & \ddots &\vdots\\
0 & \hdots & 0 & 1 & 0
\end{bmatrix}
$$
Notice that
$$
\begin{gathered}
P\Sigma_2P^{\top}=\Sigma_2 \\
P(\Sigma_1-\Sigma_2)P^{\top}=\sigma_1^2\begin{bmatrix}
\begin{array}{ccc:ccc}
    0 & 1 & -1 & 0 & \cdots & 0\\
    1 & 0 & 1 & 0 & \cdots & 0\\
    -1 & 1 & 0 & 0 & \cdots & 0\\\hdashline
    0 & 0 & 0 & 0 & \cdots & 0\\
    \vdots & \vdots & \vdots & \vdots & \ddots & \vdots \\
    0 & 0 & 0 & 0 &\cdots & 0
\end{array}
\end{bmatrix}_{n\times n}\triangleq \Delta
\end{gathered}
$$
As a result, the set of the eigenvalues of $\Sigma_1^{\frac{1}{2}}\Sigma_2^{-1}\Sigma_1^{\frac{1}{2}} - I$ is equivalent to that of $\Sigma_2^{-1}\Delta$.

Noticed that only the first three columns of $\Sigma_2^{-1}\Delta$ have non-zero entries. Therefore, $\Sigma_2^{-1}\Delta$ only has three non-zero real eigenvalues which are also the eigenvalues of the upper left $3\times 3$ block matrix, denoted as $\left[\Sigma_2^{-1}\Delta\right]_{3\times3}$. 

As $\Sigma_2$ is a symmetric circulant matrix, the eigenvalues $\{\lambda_j,j=0\dots n-1\}$ and eigenvectors $\{\mathbf{u}_j,j=0\dots n-1\}$ of $\Sigma_2^{-1}$ write as follows~\cite{gray2006toeplitz}:
$$
\begin{gathered}
\lambda_j=\frac{1}{2\sigma_1^2[1-\cos(\frac{2\pi j}{n})]+\sigma_2^2} \\
\mathbf{u}_j=\frac{1}{\sqrt{n}}\left[1,\cos(\frac{2\pi j}{n}), \cos(\frac{4\pi j}{n}), \dots, \cos(\frac{(n-1)2\pi j}{n})\right]^{\top}
\end{gathered}
$$
Let $r:= \frac{\sigma_2^2}{2\sigma_1^2}$. 
Use the above formula, we write $\left[\Sigma_2^{-1}\Delta\right]_{3\times3}$ explicitly. Denote

\begin{figure} [!tbp]
\centering  
\includegraphics[width=0.4\textwidth]{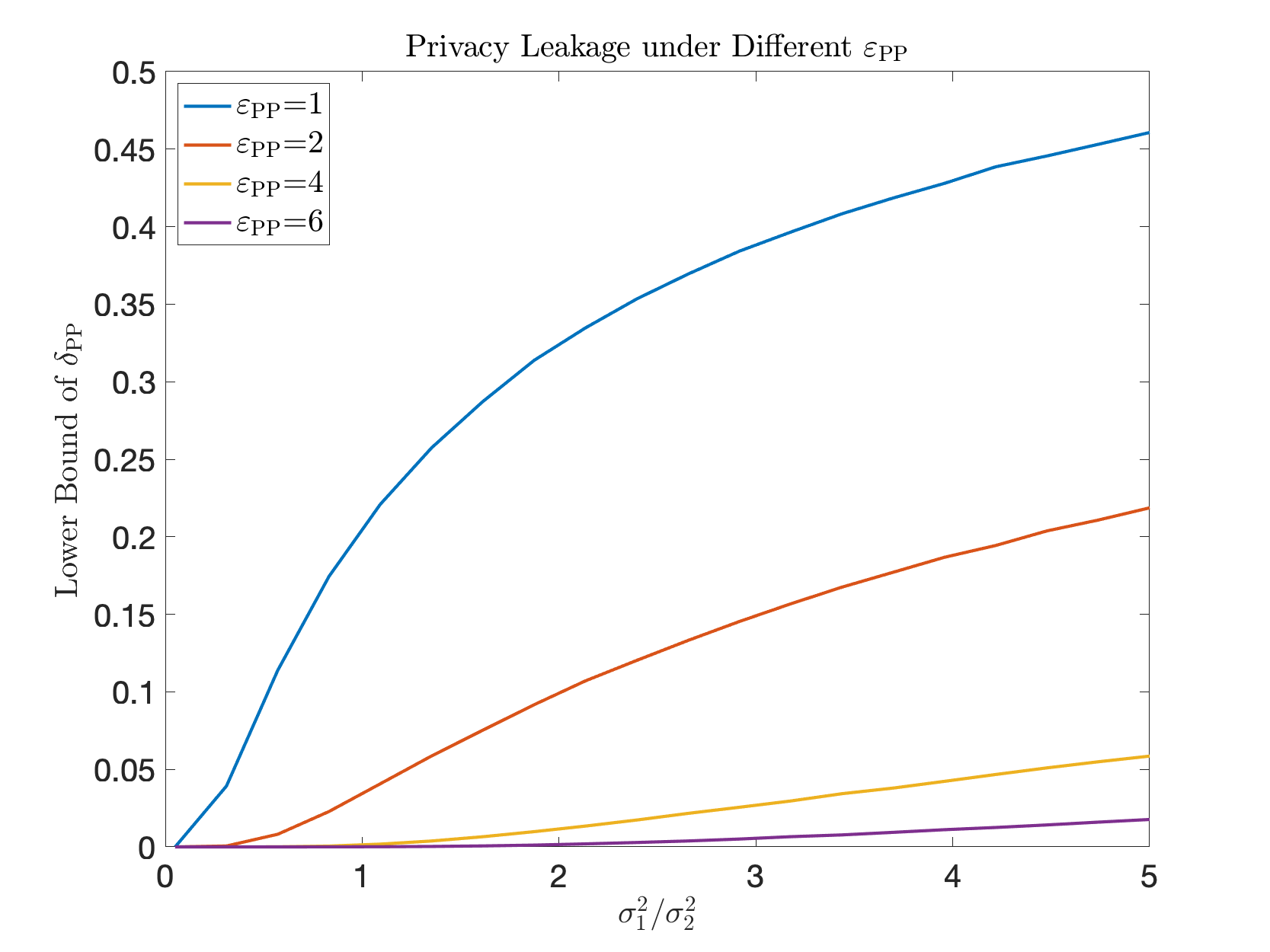}
\caption{Simulation results for \autoref{theorem:PPDP}: Greater $\sigma_1^2/\sigma_2^2$ values correlate with increased privacy leakage.}
\label{MonterCarlo}
\end{figure}

\begin{align*}
    a(r) &\triangleq \frac{1}{n}\sum_{j=0}^{n-1} \frac{\sigma_1^2}{2\sigma_1^2[1-\cos(\frac{2\pi j}{n})]+\sigma_2^2} \\
    b(r) &\triangleq \frac{1}{n}\sum_{j=0}^{n-1} \frac{\sigma_1^2\cos(\frac{2\pi j}{n})}{2\sigma_1^2[1-\cos(\frac{2\pi j}{n})]+\sigma_2^2} \\
    c(r) &\triangleq \frac{1}{n}\sum_{j=0}^{n-1} \frac{\sigma_1^2\cos(\frac{4\pi j}{n})}{2\sigma_1^2[1-\cos(\frac{2\pi j}{n})]+\sigma_2^2} \\
\end{align*}


We have
$$
\left[\Sigma_2^{-1}\Delta\right]_{3\times3}=\begin{bmatrix}
    b-c & a+c & b-a\\
    a-b & 2b  & a-b\\
    b-a & a+c & b-c
\end{bmatrix}
$$
Notice that $k_2(r)=a-c$ is an eigenvalue with eigenvector $\frac{1}{\sqrt{2}}[1,0,1]^{\top}$, and the other two satisfy
$$
\quad k_1(r)+k_3(r)=4b-a-c\quad k_1(r)k_3(r)=2(b^2-a^2+b^2-ac)
$$
We have following bound: for any measurable set $\mathcal{O} \subset \mathbb{R}^n$,
$$
\mathbf{\Pr}(b_{i1}\in\mathcal{O})\leq\delta(\varepsilon,r)+\exp(\varepsilon)\mathbf{\Pr}(b'_{i1}\in\mathcal{O})
$$
where 
$$
\begin{aligned}    
& \delta(\varepsilon,r) \\ = &\mathbf{\Pr}_{y\sim\mathcal{N}(0,I)}\left(k_1(r)y_1^2+k_2(r)y_2^2+k_3(r)y_3^2\geq 2\varepsilon-\ln2\right)
\end{aligned}
$$
Since $y_1,y_2,y_3$ are independent, $\delta(\varepsilon,r)$ can be easily obtained by a Monte Carlo simulation. 
Similarly, we can show that
$\mathcal{O} \subset \mathbb{R}^n$,
$$
\mathbf{\Pr}(b'_{i1}\in\mathcal{O})\leq\delta(\varepsilon,r)+\exp(\varepsilon)\mathbf{\Pr}(b_{i1}\in\mathcal{O})
$$

Finally, we apply Lemma~\ref{DPsequential} to obtain the $(\varepsilon_{\rm PP},\delta_{\rm PP})-$DP result for Algorithm 3, and $\varepsilon_{\rm PP}=W\varepsilon,\delta_{\rm PP}=W\delta(\varepsilon,r)$. 

We also include numerical simulations with $W=1$ in \autoref{MonterCarlo} to confirm our intuition that larger values of $\sigma_1^2/\sigma_2^2$ result in increased privacy leakage.
When $\sigma_1^2/\sigma_2^2$ is raised, holding $\varepsilon_{PP}$ constant, we observe a corresponding increase in the lower bound of $\delta_{PP}$, signifying heightened privacy leakage.

\hfill $\blacksquare$

\section{Proof of the Determinant Bound}
\label{proof:detbound}

By ~\cite[Theorem3.1]{gray2006toeplitz}, we have $\frac{|\Sigma_2|}{|\Sigma_1|} = f(r,n)$ where

$$
\begin{aligned}
\ln f(r,n) = & -\ln(1+r) + \sum_{i=0}^{n} \ln\left(1+r-\cos(\frac{2\pi i}{n})\right) \\
& - \sum_{i=0}^{n-1} \ln\left(1+r-\cos(\frac{2\pi i}{n-1})\right)
\end{aligned}
$$

\subsection{Upper Bound}
Firstly, we can assume $n$ is even.
Denote 
$$
T(n,i) := \ln\left(1+r-\cos(\frac{2\pi i}{n})\right)
$$
One has
$$
\begin{aligned}
& \ln f(r,n) \\
= & -\ln(1+r) + \sum_{i=0}^{n} T(n,i) - \sum_{i=0}^{n-1} T(n-1,i) \\
= & \ln{\frac{2+r}{1+r}} + 2 \sum_{i=0}^{\frac{n}{2}-1} (T(n,i) - T(n-1,i))\\
< & \ln{\frac{2+r}{1+r}}
\end{aligned}
$$
where the last inequality comes from
$$
\begin{gathered}
\cos(\frac{2\pi i}{n}) > \cos(\frac{2\pi i}{n-1}), i = 1, \ldots, \frac{n}{2}-1; \\
\cos(\frac{2\pi i}{n}) > \cos(\frac{2\pi (i-1)}{n-1}), i = \frac{n}{2}+1, \ldots, n-1
\end{gathered}
$$

Thus, when $n$ is even, one has $f(r,n) < 2$. A similar proof works for the case where $n$ is odd.

\subsection{Lower Bound}

We define

$$
\begin{aligned}
& \ln g(r, n)= \ln f(r, n)-\ln (2+r)+\ln (1+r) \\
& =-\ln (2+r)+\sum_{i=0}^n \ln \left(1+r-\cos \left(\frac{2 \pi i}{n}\right)\right)-\sum_{i=0}^{n-1} \ln \left(1+r-\cos \left(\frac{2 \pi i}{n-1}\right)\right) \\
\end{aligned}
$$

We have
$$
\begin{aligned}
& \frac{\partial \ln g(r, n)}{\partial r}=-\frac{1}{2+r}+\sum_{i=0}^n \frac{1}{1+r-\cos \left(\frac{2 \pi i}{n}\right)}-\sum_{i=0}^{n-1} \frac{1}{1+r-\cos \left(\frac{2 \pi i}{n-1}\right)}
\end{aligned}
$$

With
$$
\begin{aligned}
& \quad \cos \left(\frac{2 \pi i}{n}\right)>\cos \left(\frac{2 \pi i}{n-1}\right) \quad i=1, \ldots, \frac{n}{2}-1 ; \\
& \cos \left(\frac{2 \pi i}{n}\right)>\cos \left(\frac{2 \pi (i-1)}{n-1}\right), \quad i=\frac{n+1}{2}, \ldots, n-1,
\end{aligned}
$$

we get $\frac{\partial \ln g(r, n)}{\partial r} > 0$.
So $\ln g(r, n)$ is monotonically increasing with $r$.
$\forall r, \forall n$, we have

$$
\begin{aligned}
\ln g(r, n) & > \lim _{r \rightarrow 0} \ln g(r, n) \\
& =\sum_{i=1}^{n-1} \ln \left(1-\cos \left(\frac{2 \pi i}{n}\right)\right)-\sum_{i=1}^{n-2} \ln \left(1-\cos \left(\frac{2 \pi i}{n-1}\right)\right) -\ln 2
\end{aligned}
$$
Assume 
$$
h(n) = \sum_{i=1}^{n-1} \ln \left(1-\cos \left(\frac{2 \pi i}{n}\right)\right)-\sum_{i=1}^{n-2} \ln \left(1-\cos \left(\frac{2 \pi i}{n-1}\right)\right)
$$

We have
$$
\begin{aligned}
& h(n)-h(n-1) \\
 = & \sum_{i=1}^{n-1} \ln \left(2 \sin ^2 \frac{\pi i}{n}\right)-2 \sum_{i=1}^{n-2} \ln \left(2 \sin ^2 \frac{\pi i}{n-1}\right)+\sum_{i=1}^{n-3} \ln \left(2 \sin ^2 \frac{\pi i}{n-2}\right) \\
= &\sum_{i=1}^{n-1} \ln \left(\sin \frac{\pi i}{n}\right)-2 \sum_{i=1}^{n-2} \sin \frac{\pi i}{n-1}+\sum_{i=1}^{n-3} \ln \sin \frac{\pi i}{n-2} \\
= & \ln \frac{\prod_{k=1}^{n-1} \sin \frac{k \pi}{n} \cdot \prod_{k=1}^{n-3} \sin \frac{k \pi}{n-2}}{\prod_{k=1}^{n-2} \sin \frac{k \pi}{n-1} \cdot \prod_{k=1}^{n-2} \sin \frac{k \pi}{n-1}}
\end{aligned}
$$

According to $\prod_{k=1}^{n-1} \sin \frac{k \pi}{n}=\frac{n}{2^{n-1}}$, we have:

$$
\begin{aligned}
h(n)-h(n-1) =\ln \frac{n^2-2n}{n^2-2n+1} < 0
\end{aligned}
$$
So $h(n)$ is monotonically decreasing with $n$.
Then we have 

$$
\begin{aligned}
& \ln g(r, n) > \lim _{r \rightarrow 0} g(r, n) = h(n) - \ln 2 \\
> & \lim _{n \rightarrow \infty} \left( \sum_{i=1}^{n-1} \ln \left(1-\cos \left(\frac{2 \pi i}{n}\right)\right)-\sum_{i=1}^{n-2} \ln \left(1-\cos \left(\frac{2 \pi i}{n-1}\right)\right) \right) - \ln 2 \\
= & \frac{1}{2\pi} \int_{0}^{2\pi} \ln(1-\cos x)dx - \ln 2 \\
= & -2 \ln 2
\end{aligned}
$$

Therefore,
$$
\ln f(r, n)-\ln (2+r)+\ln (1+r)>-2 \ln 2
$$
and
$$
f(r+n)> \exp(-2 \ln 2+\ln \frac{2+r}{1+r}) >\frac{1}{4}
$$

\hfill $\blacksquare$

\section{Communication Overhead Analysis}
\label{appendix: communication}

Let $n, l, W$ denote the number of instances, splitting candidates, and noise vectors per splitting candidate, respectively. For each node split, the communication overhead of different methods is:

\begin{itemize}
    \item \textbf{Other DP methods:} $(2n+l)$ messages.
    \item \textbf{MaskedXGBoost:} $(lW+2l)n+l$ messages.
    \item \textbf{HE method:} $(2n+2l) \times ct$ messages, where $ct$ is the ciphertext size.
\end{itemize}

For $n=1000, l=32, W=3$, and HE method (Paillier) using 4096 bits encryption, the overhead is 16KB, 1.2MB, 2.1MB respectively. Our method has a lower overhead than the HE method but is still higher than other DP methods.

\section{Adversary Analysis}
\label{appendix: adversary}

We further provide analysis on adversaries deviating from the protocol, building upon the honest-but-curious results in \autoref{subsec:exp_attackforAP} and \autoref{subsec:exp_attackforPP}.

The protocol involves three components: the PP constructs the noise vectors (\autoref{alg:Secure-Noise}), the AP adds noise based on these vectors (\autoref{alg:Secure-Response}), and the AP calculates the splitting score to determine the optimal split(\autoref{alg:1}).

\begin{itemize}
    \item The $\sigma_2$ component of the noise vector constructed by the PP protects its own privacy (as demonstrated in \autoref{theorem:PPDP}). However, as indicated by \autoref{theorem:utility}, increasing $\sigma_2$ leads to a degradation of the model's utility. If the PP deviates from the protocol by reducing $\sigma_2$, it will expose more privacy; conversely, increasing $\sigma_2$ would compromise the goal of improving utility.
    \item For the AP, noise is added according to the parameter $C$. If the AP deviates from the protocol by reducing $C$, the privacy protection of its gradients will be weakened (as per \autoref{theorem:APLDP}). Increasing $C$ would introduce additional disturbing noise into the vector, thereby reducing the utility, as to \autoref{theorem:utility}.
    \item If the AP deviates from the protocol in \autoref{alg:1}, it will fail to identify the optimal split, leading to a decrease in the model's utility.
\end{itemize}

In our vertical federated learning setting with two parties, if either party violates the protocol, it will only affect its own privacy and will not improve its ability to launch a privacy attack against the other party. Additionally, it will reduce the model's utility.

\section{Additional Experimental Results}
\label{resultsappendix}

\begin{figure*}[!tbp]
\centering
\includegraphics[width=180mm]{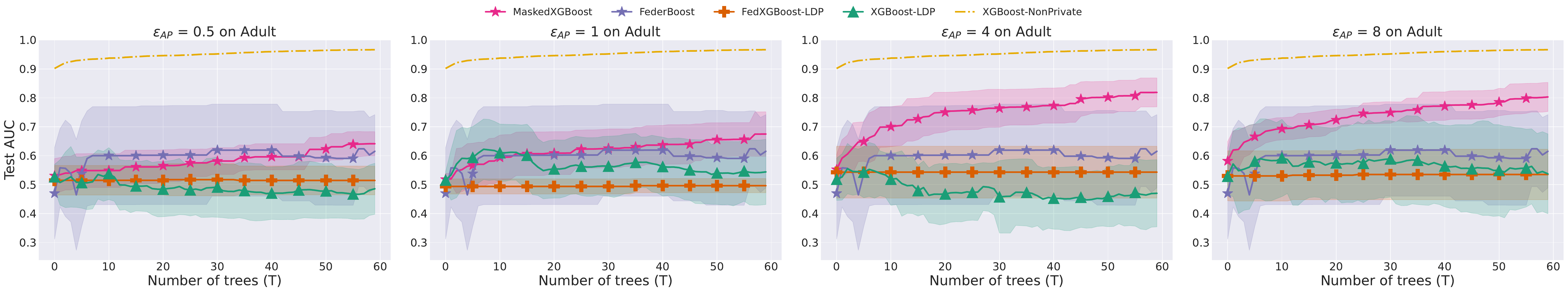}
\includegraphics[width=180mm]{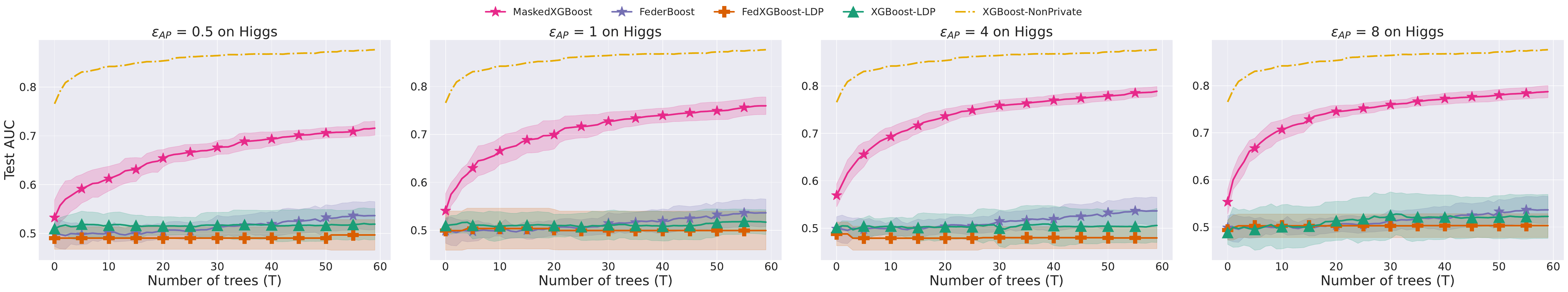}
\includegraphics[width=180mm]{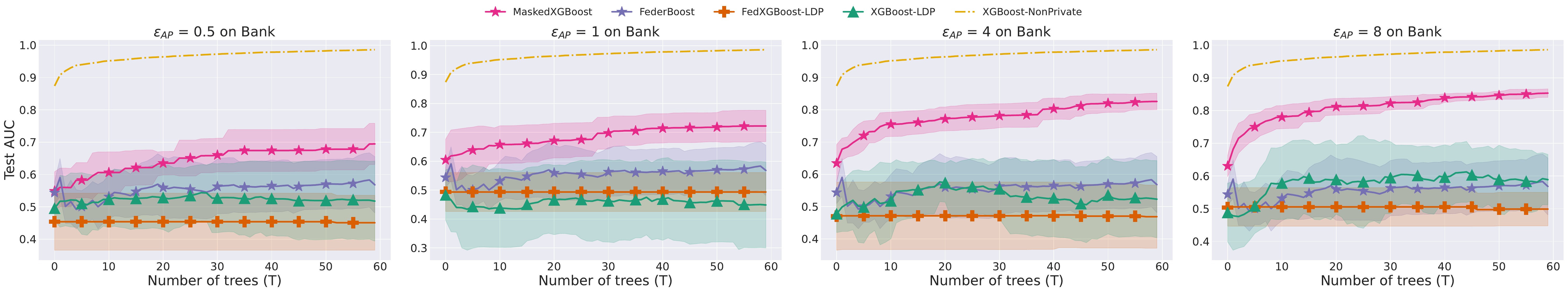}
\includegraphics[width=180mm]{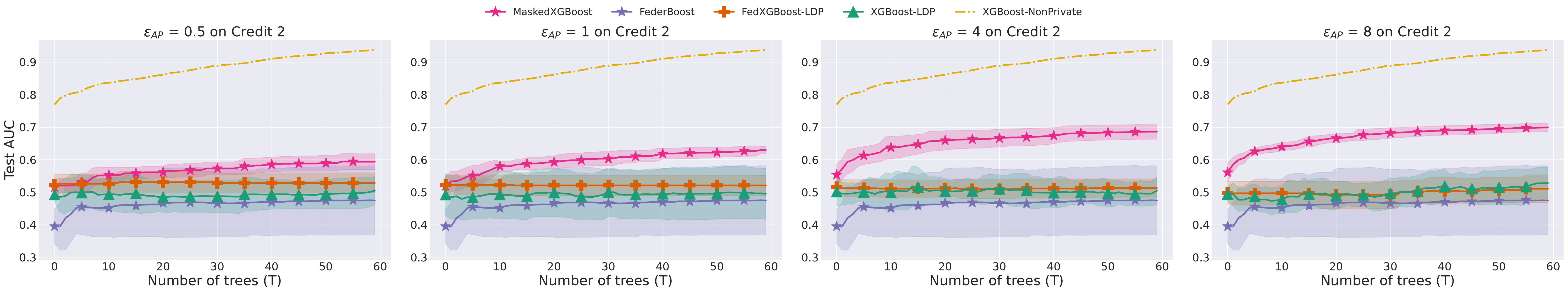}
\includegraphics[width=180mm]{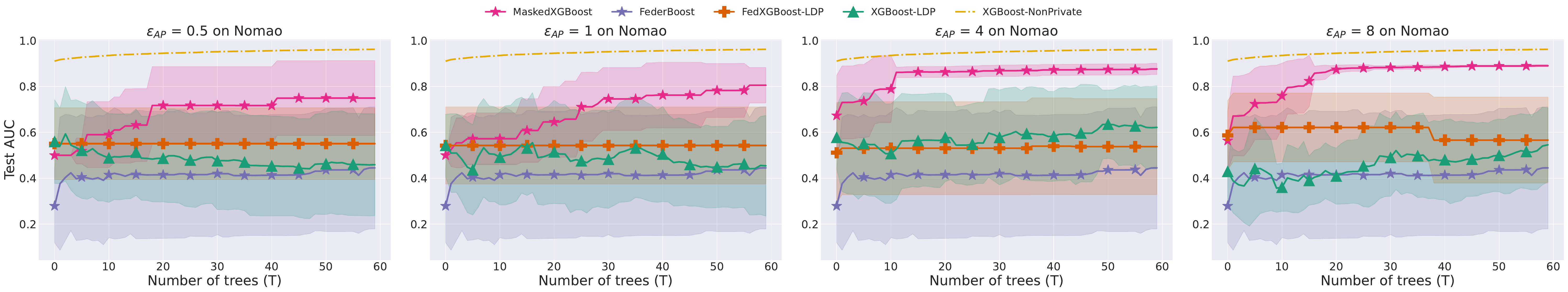}
\caption{Training process of the four XGBoost algorithms with different privacy budgets on Adult, Higgs, Bank, Credit 2, and Nomao datasets.}
\label{trainprocessepsappendix}
\end{figure*}

\begin{figure*}[!tbp]
\centering
\includegraphics[width=100mm]{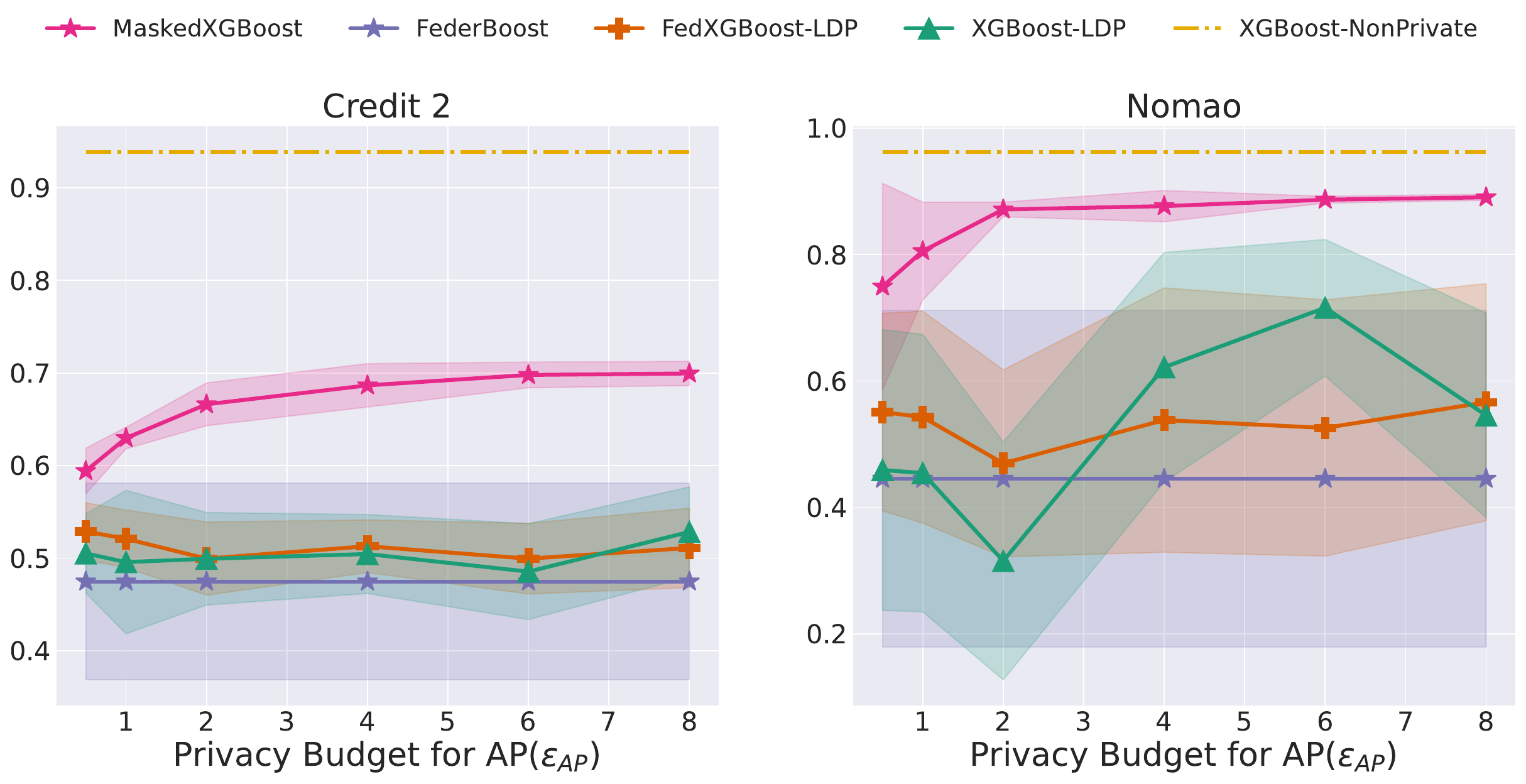}
\caption{Utility of different methods for different privacy budgets on Credit 2 and Nomao.}
\label{utility:credit2&nomao}
\end{figure*}

\begin{figure*}[!tbp]
\centering
\includegraphics[width=0.7\textwidth]{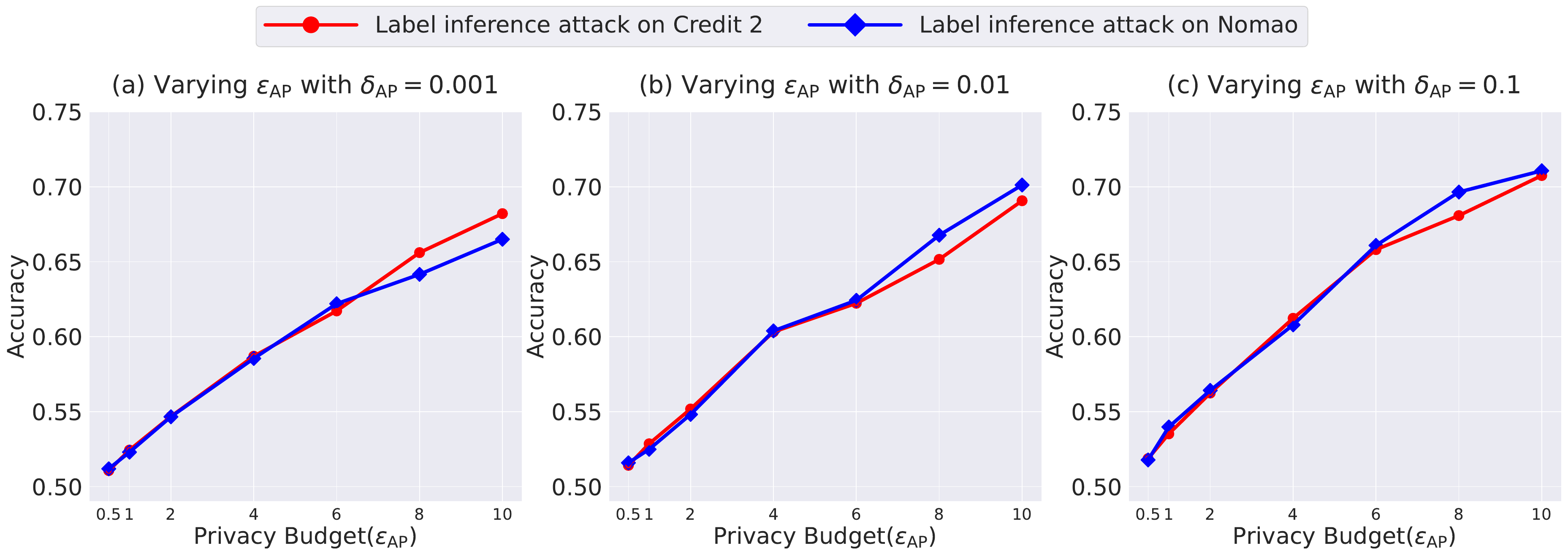}
\caption{Empirical privacy evaluation for AP with varying privacy budget $\varepsilon_{AP}$ and $\delta_{AP}$. 
This is evaluated on Credit 2 and Nomao.}
\label{1000attackappendix}
\end{figure*}

\begin{figure*}[!tbp]
\centering
\includegraphics[width=0.7\textwidth]{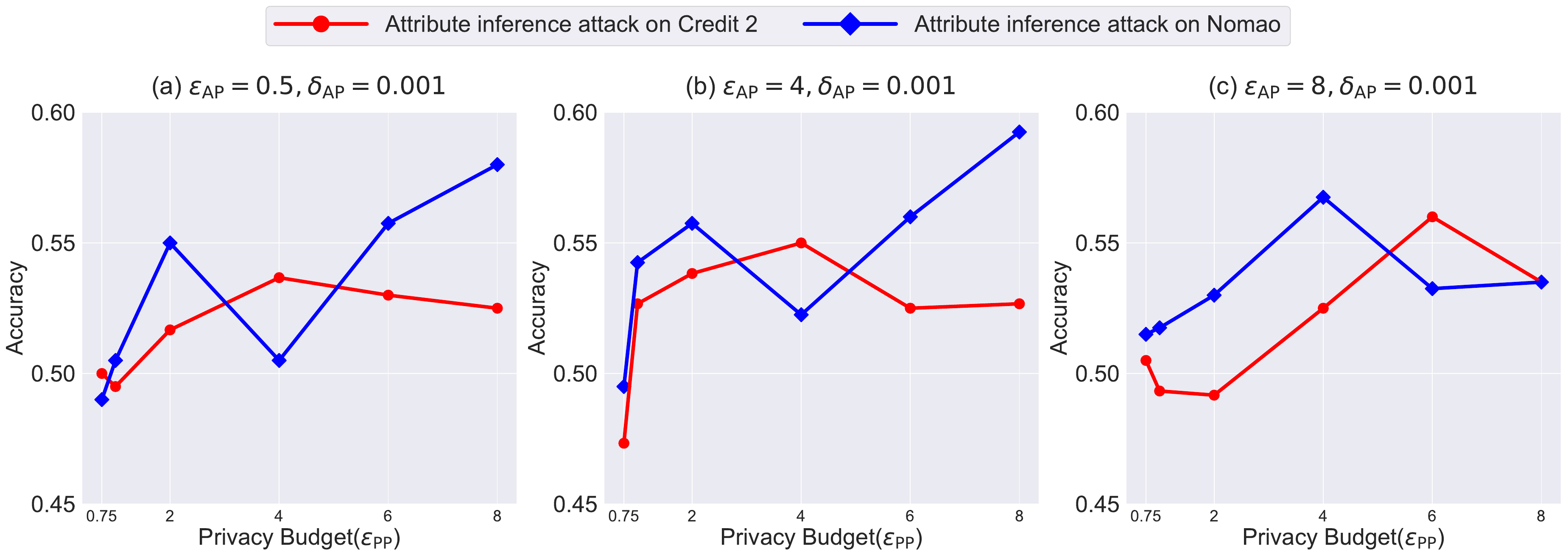}
\caption{Empirical privacy evaluation for PP with varying privacy budget $\varepsilon_{\rm PP}$ and $\varepsilon_{\rm AP}$.
This is evaluated on Credit 2 and Nomao.
}
\label{apattack_twodatasets}
\end{figure*}

\begin{figure*}[!tbp]
\centering
\includegraphics[width=180mm]{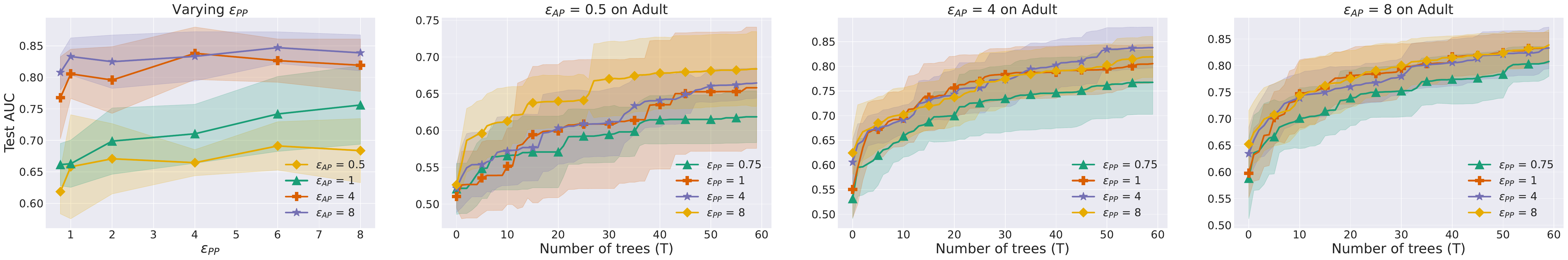}
\includegraphics[width=180mm]{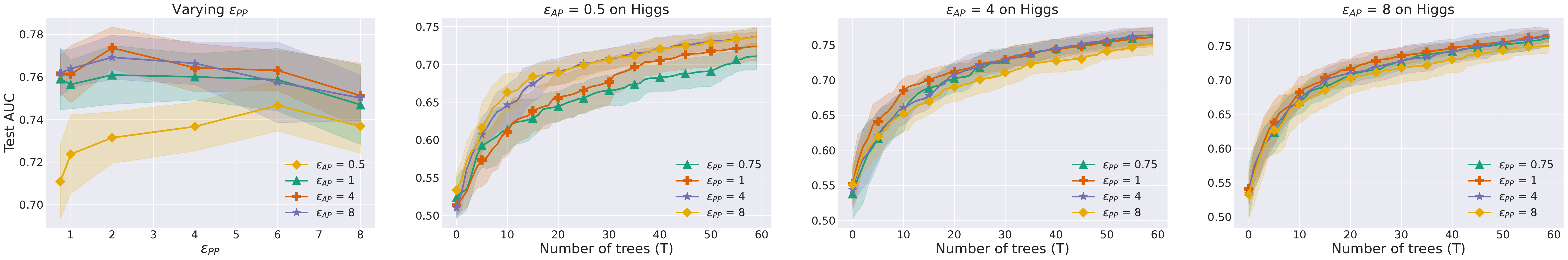}
\includegraphics[width=180mm]{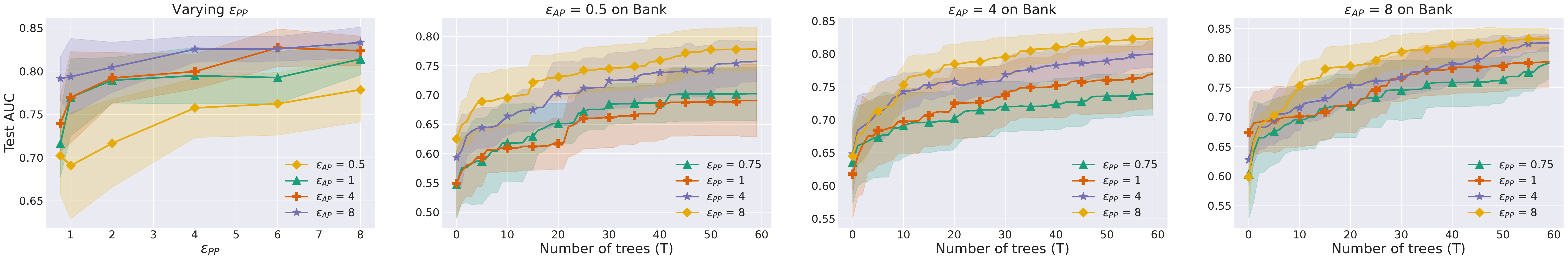}
\includegraphics[width=180mm]{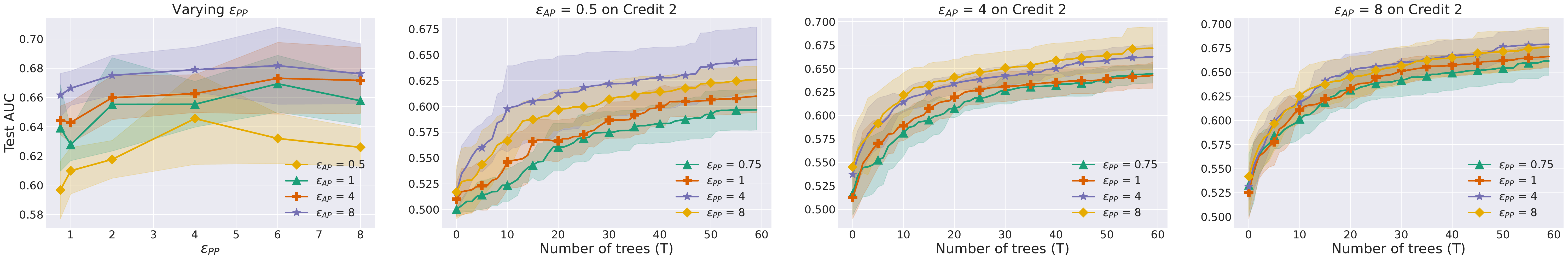}
\includegraphics[width=180mm]{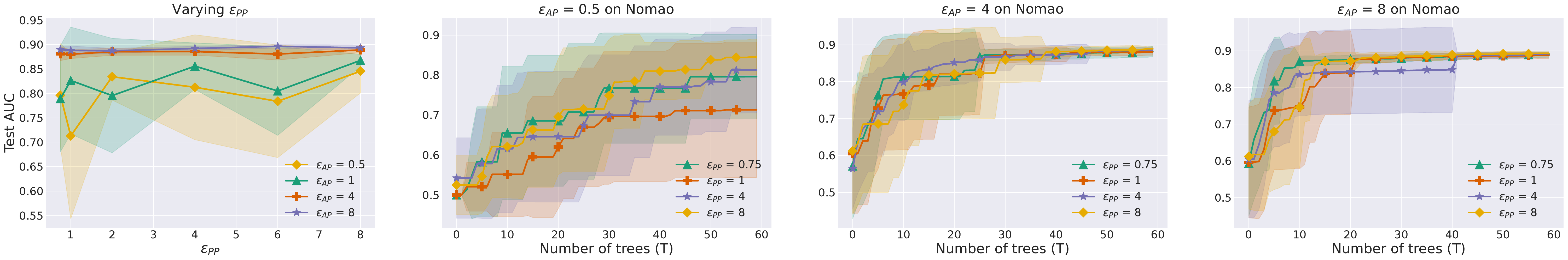}
\caption{Ablation study on Adult, Higgs, Bank, Credit 2, and Bank datasets: the effect of $\varepsilon_{PP}$ on final test AUC and training process.}
\label{ablationappendix}
\end{figure*}

\autoref{utility:credit2&nomao} illustrates the utility evaluation results on Credit 2 and Nomao datasets.
\autoref{trainprocessepsappendix} is the convergence evaluation results on Adult, Higgs, Bank, Credit 2, and Nomao datasets.
\autoref{1000attackappendix} is the empirical privacy evaluation results for AP on Credit 2 and Nomao datasets.
\autoref{apattack_twodatasets} is the empirical privacy evaluation results for PP on Credit 2 and Nomao datasets.
\autoref{ablationappendix} shows the ablation study results on Adult, Higgs, Bank, Credit 2, and Nomao datasets.

\end{document}